\documentclass[conference]{IEEEtran}

\AtBeginDocument{%
  \providecommand\BibTeX{{%
    \normalfont B\kern-0.5em{\scshape i\kern-0.25em b}\kern-0.8em\TeX}}}

\usepackage{tikz}
\usepackage{mathtools}
\usepackage{amsfonts, amsthm} 
\usepackage{algorithm}
\usepackage[noend]{algpseudocode}
\usepackage[caption=false,font=scriptsize]{subfig}
\usepackage{balance}
\usepackage[hidelinks]{hyperref}
\usepackage{filecontents}
\DeclareMathOperator*{\argmin}{arg\,min}
\DeclareMathOperator*{\argmax}{arg\,max}

\usepackage{enumitem}
\usepackage{etoolbox}
\usepackage{tikz}
\usetikzlibrary{tikzmark}
\usetikzlibrary{calc}
\errorcontextlines\maxdimen

\newcommand{\ALGtikzmarkcolor}{black}% customise this, if you want
\newcommand{\ALGtikzmarkextraindent}{4pt}% customise this, if you want
\newcommand{\ALGtikzmarkverticaloffsetstart}{-.5ex}% customise this, if you want
\newcommand{\ALGtikzmarkverticaloffsetend}{-.5ex}% customise this, if you want
\makeatletter
\newcounter{ALG@tikzmark@tempcnta}

\newcommand\ALG@tikzmark@start{%
    \global\let\ALG@tikzmark@last\ALG@tikzmark@starttext%
    \expandafter\edef\csname ALG@tikzmark@\theALG@nested\endcsname{\theALG@tikzmark@tempcnta}%
    \tikzmark{ALG@tikzmark@start@\csname ALG@tikzmark@\theALG@nested\endcsname}%
    \addtocounter{ALG@tikzmark@tempcnta}{1}%
}

\def\ALG@tikzmark@starttext{start}
\newcommand\ALG@tikzmark@end{%
    \ifx\ALG@tikzmark@last\ALG@tikzmark@starttext
    \else
        \tikzmark{ALG@tikzmark@end@\csname ALG@tikzmark@\theALG@nested\endcsname}%
        \tikz[overlay,remember picture] \draw[\ALGtikzmarkcolor] let \p{S}=($(pic cs:ALG@tikzmark@start@\csname ALG@tikzmark@\theALG@nested\endcsname)+(\ALGtikzmarkextraindent,\ALGtikzmarkverticaloffsetstart)$), \p{E}=($(pic cs:ALG@tikzmark@end@\csname ALG@tikzmark@\theALG@nested\endcsname)+(\ALGtikzmarkextraindent,\ALGtikzmarkverticaloffsetend)$) in (\x{S},\y{S})--(\x{S},\y{E});%
    \fi
    \gdef\ALG@tikzmark@last{end}%
}

\apptocmd{\ALG@beginblock}{\ALG@tikzmark@start}{}{\errmessage{failed to patch}}
\pretocmd{\ALG@endblock}{\ALG@tikzmark@end}{}{\errmessage{failed to patch}}
\makeatother
\begin{document}
\title{Out-of-Core Edge Partitioning at Linear Run-Time}

\author{\IEEEauthorblockN{Ruben Mayer}
\IEEEauthorblockA{\textit{Technical University of Munich}\\
ruben.mayer@tum.de}
\and
\IEEEauthorblockN{Kamil Orujzade}
\IEEEauthorblockA{\textit{Technical University of Munich} \\
kamil.orujzade@tum.de}
\and
\IEEEauthorblockN{Hans-Arno Jacobsen}
\IEEEauthorblockA{\textit{University of Toronto} \\
jacobsen@eecg.toronto.edu}
}

%%%
%%% The "author" command and its associated commands are used to define the authors and their affiliations.
%\author{Ruben Mayer}
%\affiliation{%
%  \institution{Technical University of Munich}
%  %\streetaddress{P.O. Box 1212}
%  %\city{Dublin}
%  %\state{Ireland}
%  %\postcode{43017-6221}
%}
%\email{ruben.mayer@tum.de}
%
%\author{Kamil Orujzade}
%%\orcid{0000-0002-1825-0097}
%\affiliation{%
%  \institution{Technical University of Munich}
%}
%\email{kamil.orujzade@tum.de}
%
%\author{Hans-Arno Jacobsen}
%%\orcid{0000-0001-5109-3700}
%\affiliation{%
%  \institution{University of Toronto}
%}
%\email{jacobsen@eecg.toronto.edu}

\maketitle

\begin{tikzpicture}
\begin{scope}[overlay]
\node[text width=19.5cm] at ([yshift=-20.5cm,xshift=-2cm]current page.south) {(c) 2022 IEEE. Personal use of this material is permitted. Permission from IEEE must be obtained for all other uses, in any current or future media, including reprinting/republishing this material for advertising or promotional purposes, creating new collective works, for resale or redistribution to servers or lists, or reuse of any copyrighted component of this work in other works. \newline The definitive version is published in Proceedings of the 2022 IEEE 38th International Conference on Data Engineering (ICDE '22).};
\end{scope}
\end{tikzpicture}

\begin{abstract}
Graph edge partitioning is an important preprocessing step to optimize distributed computing jobs on graph-structured data. The edge set of a given graph is split into $k$ equally-sized partitions, such that the replication of vertices across partitions is minimized. Out-of-core edge partitioning algorithms are able to tackle the problem with low memory overhead. Exsisting out-of-core algorithms mainly work in a streaming manner and can be grouped into two types. While \emph{stateless} streaming edge partitioning is fast and yields low partitioning quality, \emph{stateful} streaming edge partitioning yields better quality, but is expensive, as it requires a scoring function to be evaluated for every edge on every partition, leading to a time complexity of $\mathcal{O}(|E|*k)$. In this paper, we propose 2PS-L, a novel out-of-core edge partitioning algorithm that builds upon the stateful streaming model, but achieves linear run-time (i.e., $\mathcal{O}(|E|)$). 2PS-L consists of two phases. In the first phase, vertices are separated into clusters by a lightweight streaming clustering algorithm. In the second phase, the graph is re-streamed and vertex clustering from the first phase is exploited to reduce the search space of graph partitioning to only two target partitions for every edge. Our evaluations show that 2PS-L can achieve better partitioning quality than existing stateful streaming edge partitioners while having a much lower run-time. As a consequence, the total run-time of partitioning and subsequent distributed graph processing can be significantly reduced.
\end{abstract}

%-------------------------------------------------------------------------------
\section{Introduction}
\label{sec:introduction}
In the recent past, many distributed data analytics systems have emerged that process \emph{graph-structured data}. Among the most prominent examples are distributed graph processing systems such as Spark/GraphX~\cite{graphx}, PowerGraph~\cite{powergraph}, Pregel~\cite{pregel}, and Giraph~\cite{giraph}, iterative machine learning systems such as GraphLab~\cite{10.14778/2212351.2212354}, distributed graph databases like Neo4j~\cite{10.1145/2384716.2384777}, and, more recently, distributed graph neural network (GNN) processing frameworks like DGL~\cite{dgl} and ROC~\cite{MLSYS2020_fe9fc289}. These systems commonly apply \emph{graph partitioning} to cut the graph data into partitions. By minimizing the cut size, the communication overhead in distributed computations is minimized, as each cut through the graph induces message transfer between the compute nodes while performing distributed computations.

There are two major partitioning strategies. \emph{Vertex partitioning} separates the vertices of the graph by cutting through the edges, while \emph{edge partitioning} separates the edges of the graph by cutting through the vertices. In edge partitioning, a vertex cut leads to \emph{vertex replication}, i.e., the vertex is adjacent to edges of multiple different partitions. When the distribution of vertex degrees in a graph is highly skewed (e.g., in so-called ``power-law graphs''), it has been shown that edge partitioning is more effective than vertex partitioning in finding good cuts~\cite{Bourse:2014:BGE:2623330.2623660}. The edge partitioning problem is known to be NP-hard, and hence, can only be solved heuristically for large graphs~\cite{Zhang:2017:GEP:3097983.3098033}.

Existing approaches to solve the edge partitioning problem can be categorized into two groups. \emph{In-memory} partitioners~\cite{Karypis:1998:FHQ:305219.305248, schlag2019scalable, Margo:2015:SDG:2824032.2824046, Zhang:2017:GEP:3097983.3098033, dne} load the complete graph into memory and then perform partitioning on it. In contrast to this, \emph{out-of-core} partitioners~\cite{Stanton:2012:SGP:2339530.2339722, dbh, Petroni:2015:HSP:2806416.2806424, 8416335, hep} forgo materializing a representation of the complete graph in memory. Instead, they commonly ingest the graph as a stream of edges, i.e., edge by edge, and assign each edge of the stream to a partition. The latter approach has the advantage that graphs can be partitioned with low memory overhead, as the edge set does not need to be completely kept in memory. This way, the monetary costs of graph partitioning can be reduced, as smaller machines can be used for partitioning very large graphs. For example, partitioning the Twitter graph into 32 partitions with the state-of-the-art in-memory partitioner NE~\cite{Zhang:2017:GEP:3097983.3098033} takes 28 GB of memory, while it only takes 2.7 GB with our proposed out-of-core partitioner 2PS-L.

There are two different kinds of streaming algorithms which are key to out-of-core partitioners: Stateless and stateful streaming edge partitioning. Stateless partitioning commonly assigns edges to partitions without taking into account previous assignments. Instead, partitioning is performed based on hashing over the IDs of the associated vertices. Contrary to this, stateful streaming edge partitioning keeps information about which vertex is already replicated on which partition. When assigning an edge $e$ to a partition, the partitioner computes a \emph{score} for each of the $k$ partitions, and then assigns $e$ to the partition that yields the best score. This score takes into account the vertex replication and can also use additional graph properties such as the vertex degrees. Stateful partitioning often yields better partitioning quality than stateless partitioning, as exploiting partitioning state and graph properties help to make better informed partitioning decisions~\cite{Pacaci:2019:EAS:3299869.3300076, Abbas:2018:SGP:3236187.3269471, verma-vldb}. However, this comes at the cost of higher time complexity. As stateless partitioning uses a constant-time hashing function to assign each edge to a partition, the time complexity is in $\mathcal{O}(|E|)$. But in stateful partitioning, the scoring function is computed for every combination of edges and partitions, yielding a time complexity of $\mathcal{O}(|E|*k)$. Thus, when $k$ is high, stateful streaming edge partitioning becomes very slow. This is problematic when designing an out-of-core edge partitioning algorithm based on streaming.

\begin{figure}
	\centering	
	\includegraphics[width=0.8\linewidth]{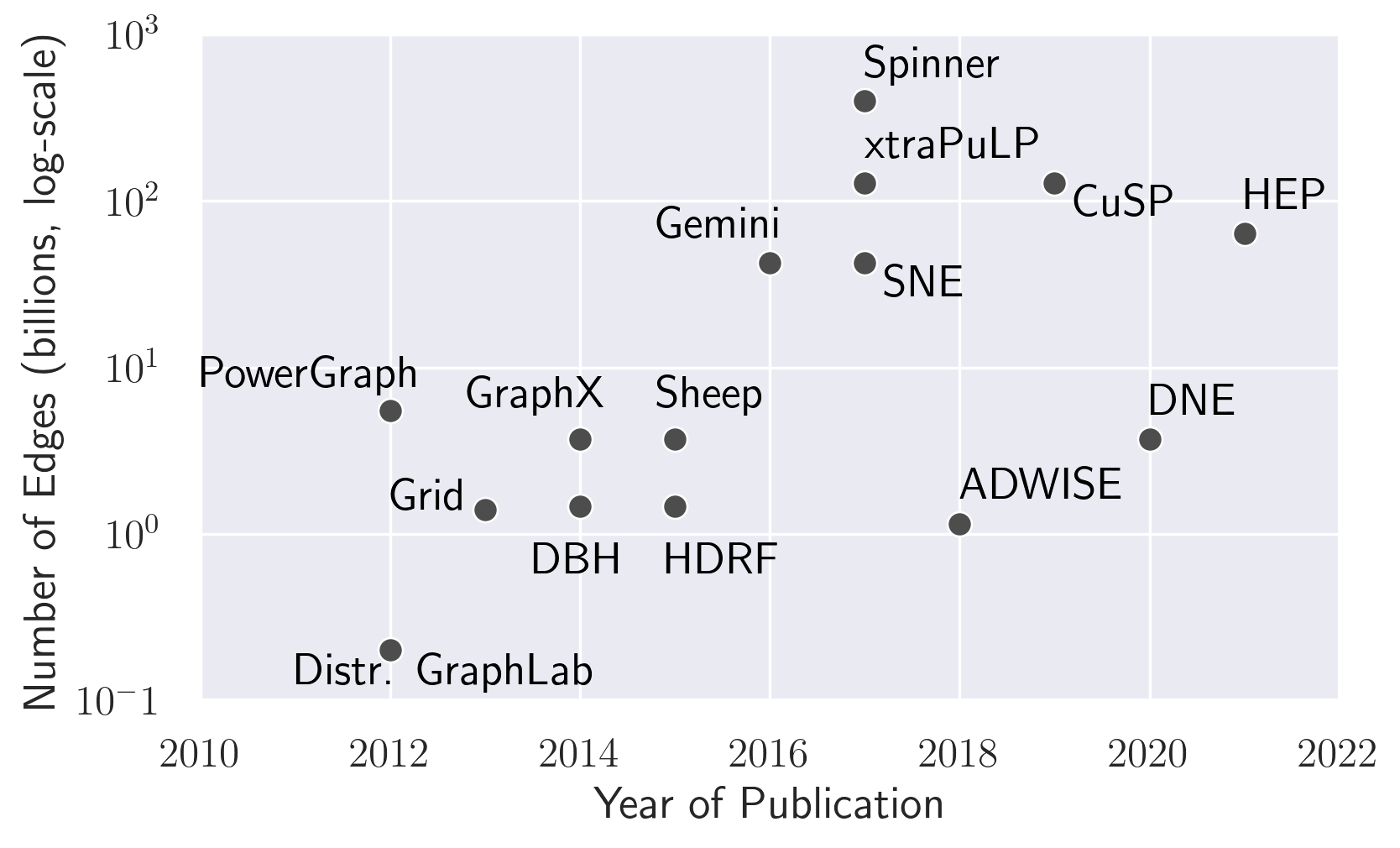}
	\vspace{-10pt}
	\caption{Size (in number of edges) of the largest real-world graph used in landmark publications on distributed graph processing and graph partitioning of the past 10 years. Synthetic graphs are not considered.}
	\label{eval:graph-sizes}
	\vspace{-8pt}
\end{figure}

Recent trends in graph processing pose the demand for partitioning graphs into an increasing number of components (i.e., require a growing value of $k$)~\cite{10.1145/3434642}. First, real-world graphs are growing in size (see Figure~\ref{eval:graph-sizes}). Second, graph computations are growing in complexity. While classical graph processing algorithms like PageRank or Connected Components involve low-complexity vertex functions such as simple aggregations of neighbors' values, emerging workloads are increasingly complex, e.g., GNN training requires for each vertex to compute a multi-layer neural network function in every iteration \cite{MLSYS2020_fe9fc289, gandhi2021p3}. These two ongoing trends make it necessary to distribute the input graph onto a growing number of compute nodes.

This development poses existing stateful streaming graph partitioning algorithms at risk of losing their practical relevance, as their run-time is becoming prohibitively large at large $k$. In practice, this problem already leads to the design of systems that forego the increased locality of high-quality graph partitioning with stateful partitioning and instead simply hash the input graph (e.g., the GNN processing framework P$^3$~\cite{gandhi2021p3}). As a consequence, the potential for efficiency gains due to higher data locality using a high-quality graph partitioner remains unexploited.

In this paper, we overcome the described limitations of stateful streaming edge partitioning by proposing a novel out-of-core algorithm that has linear run-time. This way, we push the scalability of out-of-core graph partitioning along a new dimension, namely, the number of partitions $k$, opening up new possibilities for graph partitioning applications in next-generation graph data management systems. We achieve this by combining a preprocessing phase, in which the vertices are clustered, with a streaming phase in which vertex clustering is exploited to reduce the search space of edge partitioning. Our contributions are as follows:

\begin{enumerate}[noitemsep]
	\item We propose 2PS-L, a novel two-phase out-of-core algorithm for edge partitioning. In the first phase, 2PS-L gathers information about the global graph structure by performing streaming clustering. In the second phase, clustering information is exploited to perform high-quality edge partitioning. This way, we exploit the flexibility of graph clustering and at the same time solve the more rigid edge partitioning problem. 
	\item We propose a \emph{new constant-time scoring mechanism} for streaming edge partitioning that only computes a score for \emph{two} partitions---independent of the number of partitions. As part of this new mechanism, we introduce a \emph{new scoring function} that not only takes into account vertex degrees and replication state, but also the volume of vertex clusters. As a result, the partitioning quality achieved with 2PS-L is in most cases better than existing stateful streaming edge partitioners, at a much lower run-time. \emph{Our approach introduces a novel stateful streaming mechanism that does \underline{not} involve computing a scoring function for every edge on every partition.}	
	\item We provide a thorough theoretical analysis of 2PS-L regarding time and space complexity. We show that time complexity of 2PS-L is linear in the number of edges, and space complexity is linear in the number of vertices and partitions. Hence, 2PS-L has a better time complexity than existing stateful streaming edge partitioners, while the space complexity remains the same. 2PS-L is among the first out-of-core edge partitioning algorithms that has a linear run-time in the number of edges, independent of the number of partitions. Therefore, it is particularly efficient when the number of partitions is high, where state-of-the-art out-of-core (streaming) partitioning algorithms would suffer from extremely high run-time. This way, 2PS-L can achieve a lower end-to-end run-time than existing out-of-core edge partitioners. Hence, 2PS-L closes an important research gap that had so far been overlooked.
	\item We perform extensive evaluations on large real-world graphs, showing the favorable performance of 2PS-L compared to other out-of-core and in-memory partitioners. Due to its fast run-time, 2PS-L can reduce the total run-time of partitioning and subsequent distributed graph processing jobs significantly compared to other partitioners.
\end{enumerate}

The rest of the paper is organized as follows. In Section~\ref{sec:background}, we formalize the problem of edge partitioning. In Section~\ref{sec:approach}, we introduce the 2PS-L out-of-core edge partitioning algorithm. Section~\ref{sec:analysis} contains a thorough theoretical analysis of 2PS-L. In Section~\ref{sec:evaluations}, we perform extensive evaluations of 2PS-L on a variety of real-world graphs. Finally, in Section~\ref{sec:related}, we discuss related work and then conclude the paper in Section~\ref{sec:conclusions}.
%-------------------------------------------------------------------------------
%-------------------------------------------------------------------------------
\section{Problem Analysis}
\label{sec:background}

\subsection{Edge Partitioning Problem}

\emph{Formalization.} The problem of \emph{edge partitioning} is commonly specified as follows (cf. also~\cite{Zhang:2017:GEP:3097983.3098033, Bourse:2014:BGE:2623330.2623660}). The graph $G = (V, E)$ is undirected or directed and consists of a set of vertices $V$ and a set of edges $E \subseteq V \times V$. Now, $E$ shall be split into  $k>1, k \in \mathbb{N}$ partitions $P = \{p_1, ..., p_k\}$  such that $\bigcup_{i=1,...,k} p_i = E$ and  $p_i \cap p_j = \emptyset, i \neq j$, while a balancing constraint is met: $\forall p_i \in P : |p_i| \leq \alpha  * \frac{|E|}{k} $ for a given $\alpha \geq 1, \alpha \in \mathbb{R}$. The balancing constraint ensures that the largest partition does not exceed the expected number of edges multiplied by an imbalance factor $\alpha$ that limits the acceptable imbalance. We define $V(p_i)=\{x \in V | \exists y \in V : (x,y) \in p_i \lor (y,x) \in p_i \}$ as the set of vertices covered by a partition $p_i \in P$, i.e., the set of vertices that are adjacent to an edge in $p_i$. The optimization objective of edge partitioning is to minimize the \emph{replication factor} RF($p_1, \dots, p_k$)$\ = \frac{1}{|V|} \sum_{i=1,...,k}{|V(p_i)|}$.%\footnote{``Communication volume''~\cite{Margo:2015:SDG:2824032.2824046} is an alternative metric that can directly be computed from the replication factor by subtracting 1.}

\emph{Interpretation.} The replication of a vertex on multiple partitions induces synchronization overhead in distributed processing. %In particular, vertex state needs to be synchronized between distributed compute nodes that hold different partitions. 
The lower the replication factor, the lower is the synchronization overhead. This has positive effects on the performance of distributed computations. For instance, numerous studies~\cite{Zhang:2017:GEP:3097983.3098033, dne, Pacaci:2019:EAS:3299869.3300076} show that there is a direct correlation between replication factor in edge partitioning and run-time of distributed graph processing.

\subsection{Streaming Edge Partitioning}

Streaming is the dominant way of performing out-of-core edge partitioning. In particular, the space complexity of streaming partitioning is independent of the number of edges in the graph. To do so, the graph is ingested edge by edge (one edge at a time), and each edge is immediately assigned to a partition. Depending on how the edge assignment is computed, we can differentiate between stateless and stateful streaming edge partitioning.

\emph{Stateless.} In stateless streaming edge partitioning, the assignment decision of a given edge $e$ is performed independently of the assignment of other edges. This is commonly achieved by \emph{hashing} on the vertex IDs of the adjacent vertices of $e$. As a practical example, degree-based hashing (DBH)~\cite{dbh} computes a hash on the vertex ID of the vertex that has the lower degree.

\emph{Stateful.} In stateful streaming edge partitioning, the assignment of an edge to a partition is performed based on a \emph{scoring function} that is computed for every partition. The scoring function can take into account \emph{graph properties} (e.g., the known or estimated degrees of the adjacent vertices of the edge~\cite{dbh, Petroni:2015:HSP:2806416.2806424}) as well as \emph{partitioning state} (e.g., the vertex cover sets of the partitions and the current size of the partitions~\cite{Petroni:2015:HSP:2806416.2806424}). The size of the partitioning state is limited to $\mathcal{O}(|V|)$, i.e., to only vertex-related information. As a practical example, HDRF~\cite{Petroni:2015:HSP:2806416.2806424} is a streaming partitioner that assigns an edge $e =(u,v)$ to the partition $p$ which maximizes a scoring function $C^{\mathit{HDRF}}(u,v,p) = C_{\mathit{REP}}(u,v,p) + C_{\mathit{BAL}}(p)$, where $C_{\mathit{REP}}(u,v,p)$ is a degree-weighted replication score and $C_{\mathit{BAL}}(p)$ is a balancing score. $C_{\mathit{REP}}(u,v,p)$ is highest if both vertices $u$ and $v$ adjacent to an edge $e$ are in the vertex cover set of the same partition~$p$; $C_{\mathit{BAL}}(p)$  is highest when $p$ contains the least number of edges.

\begin{figure}
	\centering	
	\subfloat[Replication facor.]{\label{a}   \includegraphics[width=0.49\linewidth]{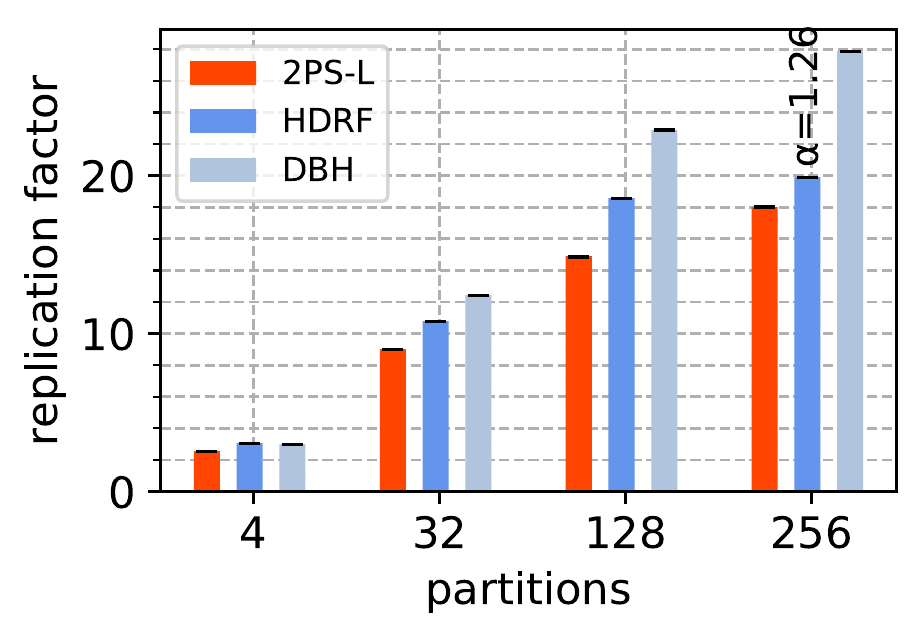}}
	\subfloat[Run-time.]{\label{b}   \includegraphics[width=0.49\linewidth]{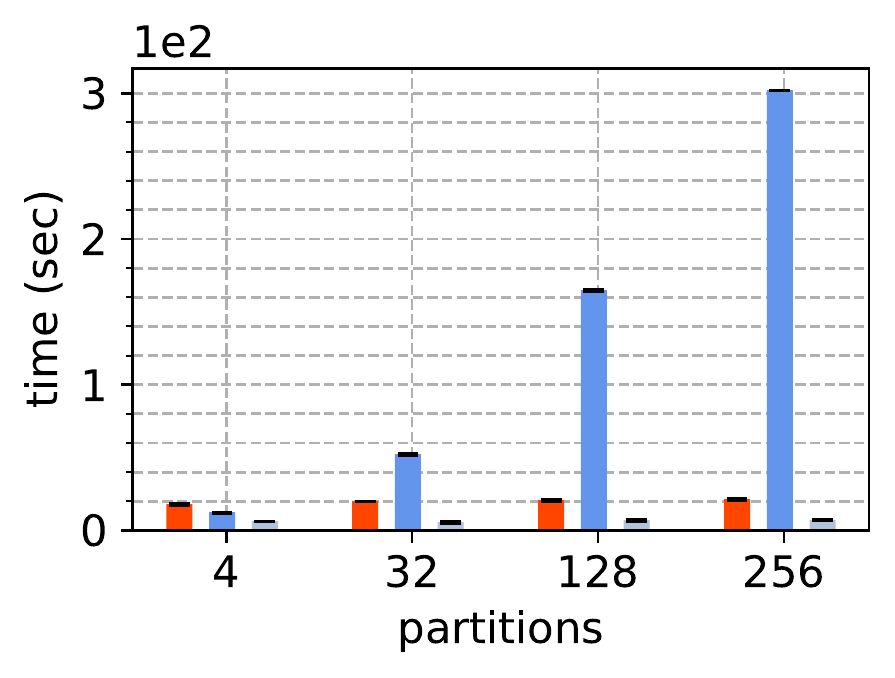}} 
	%\vspace{-5pt}
	\caption{Replication factor and run-time of 2PS-L against stateful (HDRF) and stateless (DBH) streaming partitioning on OK graph (cf. Table~\ref{tab:graphs}) at different numbers of partitions.}
	\label{eval:motivation}
	\vspace{-5pt}
\end{figure}

\paragraph*{Discussion} Stateful partitioning yields in most cases a lower replication factor than stateless partitioning~\cite{verma-vldb, Abbas:2018:SGP:3236187.3269471, Pacaci:2019:EAS:3299869.3300076}. However, as the run-time of stateful partitioning increases linearly with the number of partitions, for a growing number of partitions, it becomes less and less profitable to perform stateful partitioning. The core problem of stateful streaming edge partitioning is that the scoring function is computed for every edge on \emph{every} partition, making it inefficient at high values of $k$. We argue that if we gather information on the graph structure in a preprocessing step, we can reduce the search space of stateful streaming edge partitioning from all $k$ partitions to only two partitions (regardless of $k$). With our novel partitioning algorithm 2PS-L, we thus achieve linear run-time at competitive partitioning quality.

In Figure~\ref{eval:motivation}, we perform streaming edge partitioning with DBH (a representative of stateless partitioning) and HDRF (stateful partitioning) at a growing number of partitions $k$. While the replication factor achieved with HDRF is better than with DBH, the run-time overhead increases considerably: At $k=256$, HDRF takes more than 5 minutes to partition the OK graph, while DBH is done in 7 seconds. In many cases, investing so much time into partitioning will not pay off. 2PS-L with its linear time complexity can build 256 partitions in 21 seconds \emph{and} achieves the lowest replication factor at the same time.

%-------------------------------------------------------------------------------
%-------------------------------------------------------------------------------
\section{Approach}
\label{sec:approach}

Our main hypothesis is that the global clustering structure of the vertices in the graph can guide stateful streaming partitioning in such a way that for a given edge, only two partitions need to be considered regardless of $k$. In particular, an edge assigment can be performed solely depending on the clusters of the two adjacent vertices. We introduce our new two-phase out-of-core algorithm 2PS-L (\textbf{2}-\textbf{P}hase \textbf{S}treaming in \textbf{L}inear runtime). It consists of (1) a streaming clustering phase, where vertices are assigned to clusters based on their neighborhood relationships, and (2) a streaming edge partitioning phase, where vertex clusters are exploited to achieve a low replication factor at linear run-time.

\subsection{Phase 1: Clustering}
\label{sec:phase1}

\subsubsection{Intuition}
We observe that in edge partitioning, a group of vertices should be replicated on the same partition if there are many edges between vertices of that group, i.e., the group is densely connected. This way, many edges can be assigned to a partition while only few vertices are added to the vertex cover set of that partition, leading to a low overall replication factor. Finding groups of vertices that are densely connected is a well-known problem called \emph{clustering} or \emph{community detection}~\cite{Newman8577, FORTUNATO201075}.

In existing streaming partitioners, e.g., HDRF~\cite{Petroni:2015:HSP:2806416.2806424} or Greedy~\cite{powergraph}, it is unknown to the partitioning algorithm whether an incoming edge is an intra-cluster edge or not, i.e., whether it is incident to vertices of the same cluster. Instead, these algorithms only consider the vertex degrees, which can be misleading, as it is not always best to cut through the highest-degree vertices. Introducing an edge buffer, as in ADWISE~\cite{8416335}, allows for ``looking into the future'' in the stream to detect clusters within that buffer. However, as shown in our evaluations (cf. Section~\ref{sec:evaluations}), the buffer-based approach fails for very large graphs, as the buffer only covers a small fraction of the complete graph and, hence, cannot detect all clusters.

\begin{figure}
\centering
  \includegraphics[width=0.95\linewidth]{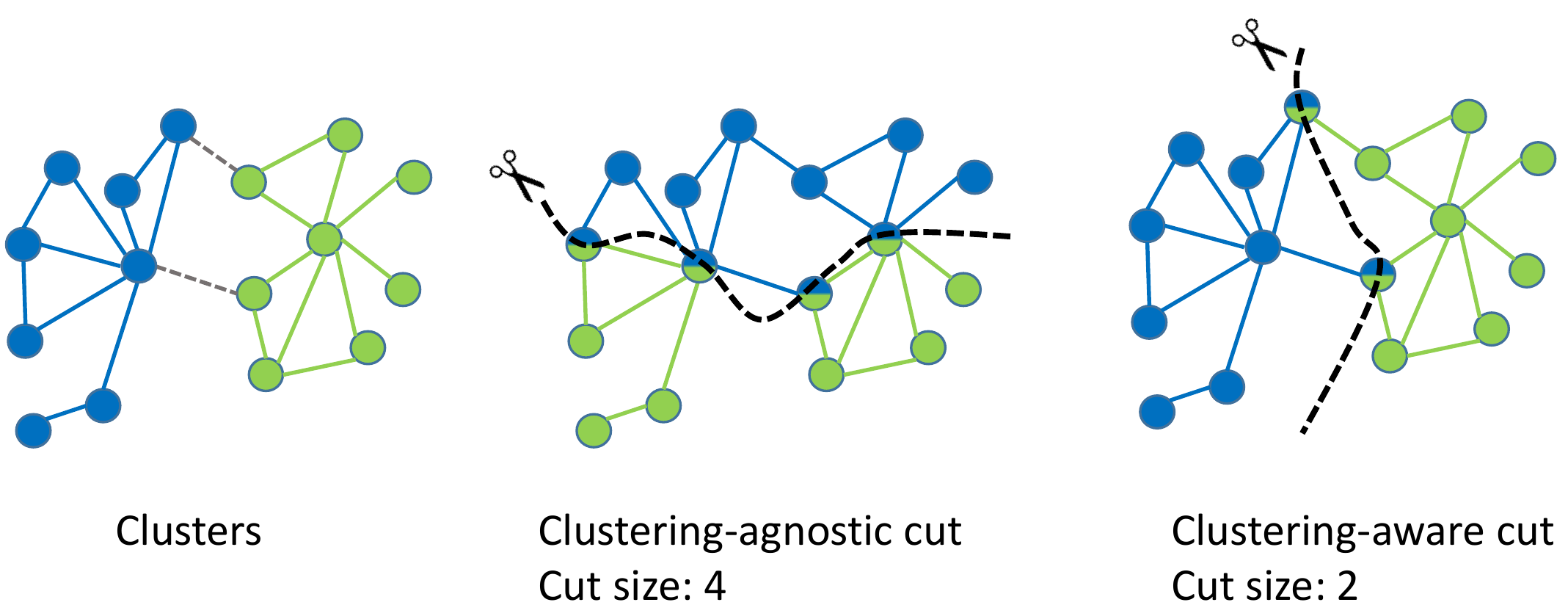}
 \vspace{-5pt}
  \caption{The awareness of graph clustering in edge partitioning leads to better partitioning quality.}
 \vspace{-5pt}
  \label{fig:clustering_idea}
\end{figure}

In 2PS-L, we go a different way: We first analyze the \emph{entire} graph in a streaming pre-processing phase to find the vertex clusters, and then exploit the global knowledge in the partitioning phase. Figure~\ref{fig:clustering_idea} illustrates this idea. In the left-hand side figure, we depict a graph structure that has two clusters, a green one and a blue one. Most of the edges are intra-cluster edges (solid lines); there are only two inter-cluster edges (dashed lines). If partitioning is performed unaware of the clustering structure of the graph, this may lead to the distribution of intra-cluster edges onto different partitions, and as a consequence, leads to a low partitioning quality. On the other hand, if partitioning is performed aware of the clustering structure of the graph, intra-cluster edges are assigned to the same partition, which leads to a high partitioning quality. 

\subsubsection{Streaming Clustering Algorithm}

In spite of similarities, clustering and edge partitioning have a different nature and, hence, are addressed with different algorithms~\cite{Newman8577}. In particular, \emph{clustering is a less constrained problem} than partitioning. First, the size of the different clusters does not have to be balanced, i.e., clusters are allowed to have different sizes (although they may have to adhere to a maximum size). Contrary to this, in edge partitioning, every partition has to cover an equal (up to the imbalance factor $\alpha$) number of edges. 
Second, the number of clusters is not necessarily predetermined, but may originate from the structure of the graph. Contrary to this, in edge partitioning, the number of partitions is determined by the user. The less constrained nature of clustering allows for devising a more efficient and flexible streaming algorithm. 

Another advantage of clustering over edge partitioning is the possibility to change the assignment of a vertex to a cluster multiple times in one single pass through the edge stream. A vertex of degree $d$ is adjacent to $d$ edges, and therefore, is visited $d$ times in one single pass through the edge stream. Every time a vertex is visited, its assignment to a cluster can be refined, taking into account new information that has been gathered since the last time the vertex was visited. Contrary to this, in edge partitioning, in a single pass through the edge list, every edge is only visited once and is immediately assigned to a partition. It is not trivially possible to revoke an edge-to-partition assignment at a later point in time when more information about the graph structure is accumulated. To re-assign edges to different partitions would require to track the mapping of \emph{edges} to partitions. Such mapping, however, can not be kept in memory for graphs with a large edge set.

\begin{algorithm}[t]
\caption{2PS-L Phase 1: Clustering}
\begin{algorithmic}[1]
\footnotesize
\State int[] \emph{d} \Comment{vertex degrees}
\State int[] \emph{vol} \Comment{cluster volumes}
\State int[] \emph{v2c} \Comment{map of vertex\_id to cluster\_id}
\State int \emph{max\_vol} \Comment{maximum cluster volume}
\State int  \emph{next\_id} $\gets$ 0 \Comment{id of next new cluster}
\vspace{0.1cm}
\Procedure{streamingClustering}{}
	\State \texttt{performStreamingPass}() 
	\State \emph{Optional: Further streaming passes.}
		
\EndProcedure

\vspace{0.1cm}
\Procedure{performStreamingPass}{}
\For{\textbf{each} $e \in $ edge\_stream}
	\For {\textbf{each} $v \in e$}
		\If{\emph{v2c}[$v$] = NULL} %\Comment{create a new cluster}
			\State \emph{v2c}[$v$] $\gets$ next\_id
			\State \emph{vol}[\emph{next\_id}] $\gets$ \emph{d}[$v$]
			\State \emph{next\_id} $\gets$ \emph{next\_id} + 1
		\EndIf
	\EndFor
	%\State $c_1 \gets $\emph{v2c}[$e$.first]; $c_2 \gets $\emph{v2c}[$e$.second]
	\If{\emph{vol}[\emph{v2c}[$v$]] $\leq$ \emph{max\_vol} $\forall v \in e$}
	\State $v_{\mathit{s}} \gets$ $v_i \in e : $ \emph{vol}[\emph{v2c}[$v_i$]] - d[$v_i$] $\leq$ \emph{vol}[\emph{v2c}[$v_j$]] - d[$v_j$]
	\State $v_{\mathit{l}} \gets$ $v_j \in e : v_j \neq v_{\mathit{s}}$ 
		\If{\emph{vol}[\emph{v2c}[$v_l$]] $+$ \emph{d}[$v_s$] $\leq$ \emph{max\_vol}}
			\State \emph{vol}[\emph{v2c}[$v_l$]] $\gets$ \emph{vol}[\emph{v2c}[$v_l$]] $+$ \emph{d}[$v_s$]]
			\State \emph{vol}[\emph{v2c}[$v_s$]] $\gets$ \emph{vol}[\emph{v2c}[$v_s$]] $-$ \emph{d}[$v_s$]]
			\State \emph{v2c}[$v_s$] $\gets$ \emph{v2c}[$v_l$] 
		\EndIf
	\EndIf
\EndFor
\EndProcedure
%\State 
\end{algorithmic}
\label{alg:clustering}
\end{algorithm}

Our streaming clustering algorithm is an extension of an algorithm by Hollocou et al.~\cite{hollocou2017streaming}. 
The intuition of Hollocou's algorithm is as follows. A given random edge from the input stream is more likely an intra-cluster edge than an inter-cluster edge---this follows directly from the understanding of a cluster as a \emph{densly connected} sub-part of the graph. Therefore, when meeting an edge $e = (u,v)$ where vertices $u$ and $v$ are currently assigned to different clusters, we draw either $u$ or $v$ into the cluster of its corresponding neighbor. We prioritize the cluster with the larger \emph{volume} (i.e., the sum of the degrees of its vertices), as a vertex is more likely to have more connections to the larger cluster than to the smaller cluster. Algorithm~\ref{alg:clustering} processes the stream edge by edge (line 10). If $u$ or $v$ have no cluster yet, we create a new cluster and assign the vertex to it (lines 11--15). Now, we compare the cluster volumes of the clusters of $u$ and $v$. The vertex that is currently assigned to the cluster with the lower volume migrates to the neighboring cluster that has the higher volume (lines 16--22). However, such migration is only allowed if the new volume of the larger cluster does not exceed a volume bound. 

Our extension introduces two novelties: \emph{bounded cluster volumes} and \emph{re-streaming}. 
(1) The original algorithm by Hollocou et al.~\cite{hollocou2017streaming} cannot guarantee that cluster volumes are bounded. This is problematic for our use case because if there are too many intra-cluster edges, we have to cut through the clusters in the subsequent partitioning phase of 2PS-L to keep the balancing constraint, which can lead to a loss of partitioning quality. Therefore, different from Hollocou et al., we compute the degree of each vertex upfront (if not already known) and use the actual vertex degree instead of the partial degree in order to compute cluster volumes. The degree of each vertex is computed in a pass through the edge set, keeping a counter for each vertex ID that is seen in an edge, which is a lightweight, linear-time operation. Furthermore, we enforce an explicit volume cap on the clusters. As we consider the actual degree of vertices instead of the partial degree, we can enforce such volume cap effectively. 
(2) Hollocou et al.~\cite{hollocou2017streaming} do not consider to apply re-streaming~\cite{Nishimura:2013:RGP:2487575.2487696} to their clustering algorithm. In re-streaming, we perform another pass through the edge list and apply exactly the same clustering algorithm, using the state from the previous pass. We evaluate the impact of the number of streaming passes on the clustering quality in our evaluations (Section~\ref{sec:restreaming}, Fig.~\ref{eval:restreaming_rf} and~\ref{eval:restreaming_runtime}).

\subsection{Phase 2: Partitioning}

The edge partitioning algorithm (Algorithm~\ref{alg:partitioning}) has three steps. First, clusters are mapped to partitions. Second, a subset of edges are pre-partitioned by exploiting vertex clustering. Third, remaining edges are partitioned by linear-time stateful streaming edge partitioning. %In the following, we describe each step in detail.

\textbf{\emph{Step 1: Mapping Clusters to Partitions.}} Our objective in the first step is to map clusters to partitions, such that the total volume of clusters across partitions is balanced. We model this problem as an instance of the classical \texttt{Makespan Scheduling Problem on Identical Machines} (\texttt{MSP-IM}). The problem can be defined as follows~\cite{graham1969bounds}:

\vspace{-3pt}
\begin{quote}Given a set of $k$ machines $M_1, ..., M_k$ and a list of $n$ jobs $j_1, ..., j_n$ with corresponding run-time $a_1,  ..., a_n$, assign each job to a machine such that the makespan (i.e., the time to complete all jobs) is minimized.\end{quote}
\vspace{-4pt}

We apply our cluster assignment problem to \texttt{MSP-IM} as follows. Partitions are corresponding to ``machines'', clusters to ``jobs'' and volumes of the clusters to ``run-times'' of the jobs. The optimization goal is to minimize the cumulative volume of the largest partition.

\texttt{MSP-IM} is NP-hard~\cite{Ullman:1975:NSP:1739944.1740138} so that we solve it by approximation. The \emph{sorted list scheduling algorithm} by Graham~\cite{graham1969bounds} is a $\frac{4}{3}$-approximation of \texttt{MSP-IM}, i.e., its result is at most $\frac{4}{3}$ times as large as the true optimum. %Sorted list scheduling works as follows: First, sort the jobs by decreasing run-time, then assign job by job from the sorted list to the least loaded machine. 
Applied to our cluster assignment problem, sorted list scheduling means that the clusters are sorted by decreasing volume (Algorithm~\ref{alg:partitioning}, line 12) and then assigned one by one to the currently least loaded partition (lines 13 to 15). %Now, it is guaranteed that the most loaded partition is at most $\frac{4}{3}$ as large as it would be in the true optimal assignment.

\textbf{Step 2: Pre-Partitioning.}
In the second step, we exploit the clustering of vertices to pre-partition a subset of edges. To do so, the pre-partitioning algorithm performs one pass through the complete edge stream (Algorithm~\ref{alg:partitioning}, line 17). For each edge $e=(e.$first$, e.$second$)$, it checks if both adjacent vertices $e$.first and $e$.second are either in the same cluster or their clusters are assigned to the same partition $p$ (cf. Step 1 discussed above). In this case, $e$ is applicable to pre-partitioning and shall be assigned to $p$ (lines 18 to 21). If $p$ is already occupied to its maximum capacity $\alpha * \frac{|E|}{k}$, $e$ is assigned to a different partition instead, using linear-time stateful streaming edge partitioning (as in Step 3).

\begin{algorithm}[t]
\caption{2PS-L Phase 2: Streaming Partitioning}
\begin{algorithmic}[1]
\footnotesize
\State int[] \emph{d} \Comment{vertex degrees (from Phase 1)}
\State int[] \emph{vol} \Comment{cluster volumes (from Phase 1)}
\State int[] \emph{v2c} \Comment{map of vertex\_id to cluster\_id (from Phase 1)}
\State int[] \emph{c2p} \Comment{map of cluster\_id to partition\_id}
\State int[] \emph{vol\_p} \Comment{sum of volumes of clusters per partition} 
\State int[][] \emph{v2p} \Comment{vertex\_id to partition\_id replication bit matrix}
\vspace{0.01cm}
\Procedure{streamingPartitioning}{}
	\State \texttt{mapClustersToPartitions}()
	\State \texttt{prepartitionEdges}() 
	\State \texttt{partitionRemainingEdges}()
\EndProcedure
\vspace{0.01cm}

\Procedure{mapClustersToPartitions}{}
\State sort clusters by volume (descending)
\For{\textbf{each} cluster $c$} (from largest to smallest)
	\State \emph{target\_p} $\gets$ $\argmin_{p_i \in P}$\emph{vol\_p}[$p_i$]	%\Comment{smallest aggregate volume}
	\State \emph{c2p}[$c$] $\gets$ \emph{target\_p}
\EndFor
\EndProcedure
\vspace{0.01cm}

\Procedure{prepartitionEdges}{}
\For{\textbf{each} $e \in $ edge\_stream}
	\State \emph{c\_1} $\gets$ \emph{v2c}[$e$.first]
	\State \emph{c\_2} $\gets$ \emph{v2c}[$e$.sec]
		\If{\emph{c\_1} = \emph{c\_2} \textbf{OR} \emph{c2p}[\emph{c\_1}] = \emph{c2p}[\emph{c\_2}]} %\Comment{both vertices are in same cluster}
		\State \emph{target\_p} $\gets$  \emph{c2p}[\emph{c\_1}]
		\If{|\emph{target\_p}| $> \alpha  * \frac{|E|}{k} $} 
			\State \emph{target\_p} is determined via scoring
		\EndIf
		\State \emph{v2p}[$e$.first][\emph{target\_p}] $\gets$ true
		\State \emph{v2p}[$e$.sec][\emph{target\_p}] $\gets$ true
		\State \texttt{output}: $e$ assigned to \emph{target\_p}
	\EndIf
\EndFor
\EndProcedure
\vspace{0.01cm}

\Procedure{partitionRemainingEdges}{}
\For{\textbf{each} $e \in $ edge\_stream}
	\State \emph{c\_1} $\gets$ \emph{v2c}[$e$.first]
	\State \emph{c\_2} $\gets$ \emph{v2c}[$e$.sec]
		\If{\emph{c\_1} = \emph{c\_2} \textbf{OR} \emph{c2p}[\emph{c\_1}] = \emph{c2p}[\emph{c\_2}]} 
		\State \textbf{continue} \Comment{skip pre-partitioned edge}
	\EndIf
	\State \emph{bestScore} $\gets 0$
	\State \emph{target\_p} $\gets$ NULL
	\For{\textbf{each} $p_i  \in \{ $\emph{c2p}[\emph{v2c}[$e$.first]], \emph{c2p}[\emph{v2c}[$e$.second]]$\}$} 
		\State \emph{score} $\gets$ $s(e.\mathit{first}, e.\mathit{second}, p_i)$ \Comment{scoring function}
		\If{\emph{score} $>$ \emph{bestScore}}
			\State \emph{bestScore} $\gets$ \emph{score}
			\State \emph{target\_p} $\gets p_i$ 
		\EndIf
	\EndFor
	\If{|\emph{target\_p}| $> \alpha  * \frac{|E|}{k} $} \Comment{degree-based hashing}
			\State \emph{target\_p} $\gets$ \texttt{hash}($\argmax_{v \in \{e.\mathit{first}, e.\mathit{second}\}}$\emph{d}[$v$])
		\EndIf
	\State \emph{v2p}[$e$.first][\emph{target\_p}] $\gets$ true
	\State \emph{v2p}[$e$.sec][\emph{target\_p}] $\gets$ true
	\State \texttt{output}: $e$ is assigned to \emph{target\_p}
\EndFor
\EndProcedure

\end{algorithmic}
\label{alg:partitioning}
\end{algorithm}

\textbf{Step 3: Streaming Partitioning.}
Edges between vertices of different clusters that are mapped to different partitions are remaining. Partitioning the remaining edges is performed with linear-time scoring-based streaming edge partitioning. We enforce a hard balancing cap, i.e., we guarantee that no partition gets more than $\alpha * \frac{|E|}{k}$ edges assigned. 

Existing stateful streaming partitioning algorithms are not aware of the vertex clustering. This induces three problems. First and foremost, the streaming algorithm has no guidance which partitions could be most suitable to place an edge on. Therefore, a scoring function is computed for \emph{every} partition. Second, the streaming algorithm starts with an empty partitioning state. Therefore, early edges in the stream are partitioned randomly at low partitioning quality. Third, the global structure of the graph is disregarded, which can lead to low partitioning quality despite of expensive scoring. 

In our linear-time scoring-based partitioning algorithm, we tackle these shortcomings as follows. First, we constrain the scoring function to only take into account two different partitions, namely, the partitions associated to the clusters of the adjacent vertices (see Step 1). This is reasonable because it is highly likely that a vertex is already replicated on the partition that is associated to its cluster; it is much less likely that a vertex is replicated on any of the other $k-1$ partitions, so that we can forego checking \emph{every} partition's state. Second, we exploit the partitioning state from pre-partitioning (see Step 2). This way, we avoid the ``cold start'' or ``uninformed assignment'' problem of streaming edge partitioning, where early edges in the stream are assigned to partitions randomly as all partitioning state is empty~\cite{8416335}. Third, in the scoring function, we take into account the cluster volumes. If a cluster has a higher volume, it is more likely that further edges that have vertices incident to the cluster will be seen in the edge stream. Thus, we assign a higher score to placing an edge on the partition that is associated with the higher-volume cluster.

\textbf{Scoring function:}
We denote with $d_v$ the degree of a vertex $v$, with $c_v$ the cluster of a vertex $v$, and with $\mathit{vol(c_v)}$ the volume of the cluster of a vertex $v$. Then, the scoring function for an edge $(u,v)$ is defined as follows:

\[ s(u,v,p) = g_u + g_v + sc_u + sc_v \]
\[ g_{\{u, v\}} =
  \begin{cases}
   1 + (1 - \frac{d_{\{u, v\}}}{d_u + d_v})       & \quad \text{if \{u,v\} is replicated on $p$} , \\
    0  & \quad \text{else.}
  \end{cases}
\] 
\[ sc_{\{u, v\}} =
  \begin{cases}
   \frac{\mathit{vol}(c_{\{u, v\}})}{\mathit{vol}(c_u) + \mathit{vol}(c_v)}       & \quad \text{if $c_{\{u, v\}}$ is assigned to $p$} , \\
    0  & \quad \text{else.}
  \end{cases}
\]

To perform streaming partitioning, 2PS-L makes a complete pass through the edge stream (Algorithm~\ref{alg:partitioning}, line 28). First, it determines whether an edge has already been pre-partitioned by checking the conditions for pre-partitioning (adjacent vertices are in the same cluster or in clusters that are mapped to the same partition). If the conditions for pre-partitioning are met, the edge is skipped (lines 29 to 33) as it has already been assigned. Else, scoring is performed on the two target partitions to determine the highest-scoring partition. If this partition has already reached its capacity bound, we hash the edge using the ID of the vertex that has the highest degree. If the hashed partition is fully occupied as well, we assign the edge to the currently least loaded partition as a last resort (not shown in the pseudo code).  

After Step 3 is finished, all edges have been assigned to partitions and none of the partitions have more than $\alpha * \frac{|E|}{k}$ edges. This concludes the 2PS-L algorithm.

%-------------------------------------------------------------------------------
%-------------------------------------------------------------------------------
\section{Theoretical Analysis}
\label{sec:analysis}
\subsection{Time Complexity}
\label{sec:time_complexity}

\renewcommand{\arraystretch}{1.25}
\begin{table}
{\footnotesize
	\begin{center}
		\begin{tabular}{|l|l|p{4.0cm}|}
			\hline
			Name & Type & Time Complexity \\	
			\hline
			2PS-L & Stateful Out-of-Core & $\mathcal{O}(|E|)$  \\ \hline \hline
			HDRF & Stateful Streaming & $\mathcal{O}(|E|*k)$  \\ \hline
			ADWISE & Stateful Streaming & $\mathcal{O}(|E|*k)$  \\ \hline \hline
			DBH & Stateless Streaming & $\mathcal{O}(|E|)$  \\ \hline 
			Grid & Stateless Streaming & $\mathcal{O}(|E|)$  \\ \hline \hline
			DNE & In-memory & $\mathcal{O}(\frac{d *|E| * (k+d)}{n * k})$ with $d = $ max. vertex degree, $n = $ num. of CPU cores \\ \hline
			METIS & In-memory & $\mathcal{O}((|V|+|E|)*\log_2(k))$ \\ \hline
			HEP & Hybrid & $\mathcal{O}(|E|*(\log{}|V| + k) + |V|)$ \\ \hline
		\end{tabular}
	\end{center}
}
%\vspace{-5pt}
\caption{Comparison of time complexity.}
\label{tab:time_complexity}
\vspace{-10pt}
\end{table}

We analyze each phase of 2PS-L separately. Phase 1, specified in Algorithm~\ref{alg:clustering}, performs a fixed number of passes through the edge set. In each pass, a constant number of operations is performed on each edge. Hence, the time complexity of the first phase is in $\mathcal{O}(|E|)$. Phase 2, specified in Algorithm~\ref{alg:partitioning}, consists of three steps. First, clusters are mapped to partitions in decreasing volume order. To sort clusters by volume is in $\mathcal{O}(|V| * \log |V|)$, as in the worst case, there are as many clusters as vertices (note that, in natural graphs, we can expect the number of clusters to be orders of magnitude smaller than the number of vertices). Each cluster is assigned to the currently least loaded partition, which can be performed in $\mathcal{O}(|V| * \log k)$ time, provided that we keep the $k$ partitions sorted by their accumulated volume while assigning clusters to them. Second, edges are pre-partitioned, such that edges whose adjacent vertices are both in clusters of the same partition are assigned to that partition. This is a constant-time operation per edge, resulting in $\mathcal{O}(|E|)$ time complexity. Third, the remaining edges are partitioned using stateful streaming, which is done in $\mathcal{O}(|E|)$ time, as for each edge, we need to compute the score against two partitions. In summary, the second phase of 2PS-L has a time complexity of $\mathcal{O}(|E|)$, as $|E| >> |V|$. The total time complexity of 2PS-L is, hence, in $\mathcal{O}(|E|)$, i.e., linear in the number of edges.

In Table~\ref{tab:time_complexity}, we compare the time complexity of 2PS-L to known results from the literature\footnote{METIS figures: http://glaros.dtc.umn.edu/gkhome/node/419}. 2PS-L is the only stateful out-of-core edge partitioner that has linear time complexity.

\renewcommand{\arraystretch}{1.25}
\begin{table}
{\footnotesize
	\begin{center}
		\begin{tabular}{|l|l|p{4.0cm}|}
			\hline
			Name & Type & Space Complexity \\	
			\hline
			2PS-L & Stateful Out-of-Core & $\mathcal{O}(|V|*k)$  \\ \hline\hline
			HDRF & Stateful Streaming & $\mathcal{O}(|V|*k)$  \\ \hline
			ADWISE & Stateful Streaming & $\mathcal{O}(|V|*k + b)$  with $b= $ buffer size  \\ \hline \hline
			DBH & Stateless Streaming & $\mathcal{O}(|V|)$  \\ \hline 
			Grid & Stateless Streaming & $\mathcal{O}(1)$  \\ \hline \hline
			--- & In-memory & $\geq \mathcal{O}(|E|)$ \\ \hline
		\end{tabular}
	\end{center}
}
%\vspace{-5pt}
\caption{Comparison of space complexity.}
\label{tab:space_complexity}
\vspace{-10pt}
\end{table}

\subsection{Space Complexity}
\label{sec:space complexity}

We analyze the data structures used in 2PS-L. In Algorithm~\ref{alg:clustering}, we use arrays to store the vertex degrees, cluster volumes and the mapping of vertices to clusters. Each of these data structures has a space complexity of $\mathcal{O}(|V|)$. In Algorithm~\ref{alg:partitioning}, besides these arrays, we use additional arrays to map the clusters to partitions and to keep the volumes of clusters per partition. These arrays all have a space complexity of $\mathcal{O}(|V|)$. Finally, we use a vertex-to-partition replication matrix, which has a space complexity of $\mathcal{O}(|V| * k)$. Hence, the overall 2PS-L algorithm has a space complexity of  $\mathcal{O}(|V| * k)$. In particular, the space complexity is independent of the number of edges in the graph. The preprocessing phase has no additional memory overhead in excess of the streaming partitioning phase. All data structures needed for clustering (i.e., vertex-to-cluster assignments, vertex degrees and cluster volumes) are directly used in the partitioning phase.
In Table~\ref{tab:space_complexity}, we compare the space complexity of 2PS-L to known results from the literature. 2PS-L has the same space complexity as other stateful out-of-core (streaming) partitioners. In-memory partitioners, by definition, have a space complexity that is at least linear in the number of edges.

\renewcommand{\arraystretch}{1.15}
\begin{table}
{\small
	\begin{center}
		\begin{tabular}{l|l|l|l|l}
			\hline
			Name & \textbf{$|V|$} & \textbf{$|E|$} & Size & Type \\	
			\hline
			com-orkut (OK) & 3.1 M & 117 M & 895 MiB & Social \\
			it-2004 (IT) & 41 M & 1.2 B & 9 GiB & Web \\
			twitter-2010 (TW) & 42 M & 1.5 B & 11 GiB & Social \\
			com-friendster (FR) & 66 M & 1.8 B & 14 GiB & Social \\
			uk-2007-05 (UK) & 106 M & 3.7 B & 28 GiB & Web \\
			gsh-2015 (GSH) & 988 M & 34 B & 248 GiB & Web \\
			wdc-2014 (WDC) & 1.7 B & 64 B & 478 GiB & Web \\	
			\hline
		\end{tabular}
	\end{center}
}
%\vspace{-5pt}
\caption{Real-world graph datasets. Size refers to the graph representation as binary edge list with 32-bit vertex IDs.}
\label{tab:graphs}
\vspace{-12pt}
\end{table}

%-------------------------------------------------------------------------------%-------------------------------------------------------------------------------
\section{Evaluations}
\label{sec:evaluations}

\textbf{Data Sets.} We evaluate on real-world graphs (cf. Table~\ref{tab:graphs}) that represent a good mix of social networks and web graphs of different sizes. OK is notoriously difficult to partition; IT, TW and FR~\cite{6413740, snapnets} are used as baselines in many graph partitioning papers~\cite{dne, hep, Zhang:2017:GEP:3097983.3098033, Petroni:2015:HSP:2806416.2806424, 8416335}. UK, GSH~\cite{BMSB, BRSLLP, BoVWFI} and WDC~\cite{wdc} represent very large graphs.

\noindent
\textbf{Baselines.} We compare to \textbf{eight} strong baselines that yield a \emph{low replication factor} among the streaming partitioners and among the in-memory partitioners, respectively. In the following, we discuss our choice of baselines.
 
\emph{(1) Streaming Partitioners. }
From the group of streaming partitioners, we compare to HDRF~\cite{Petroni:2015:HSP:2806416.2806424}, DBH~\cite{dbh}, SNE (a streaming version of the in-memory partitioning algorithm NE~\cite{Zhang:2017:GEP:3097983.3098033}), and ADWISE~\cite{8416335}. HDRF delivers a low replication factor at modest run-time and memory overhead. DBH is based on hashing and is the fastest streaming partitioner with the lowest memory overhead. SNE delivers a lower replication factor than HDRF, albeit at a significantly higher run-time and memory overhead. ADWISE uses a buffer to reorder the edge stream in order to achieve a better replication factor at the cost of higher run-time. 
Other streaming partitioners (Greedy~\cite{powergraph}, Grid~\cite{grid}, Perfect Difference Sets~\cite{grid}) are outperformed by our chosen baselines in terms of replication factor and are not considered in our study. 

\emph{(2) In-Memory Partitioners. }
From the group of in-memory partitioners, we compare to NE~\cite{Zhang:2017:GEP:3097983.3098033}, DNE~\cite{dne} and METIS~\cite{Karypis:1998:FHQ:305219.305248}. These are currently the partitioners that achieve the lowest replication factors (NE, METIS) and the best scalability (DNE). NE and METIS are single-threaded, while DNE is a parallel algorithm.
KaHiP~\cite{schlag2019scalable} is reported to provide very good partitioning quality, but it ran out of memory even on our large machine. Other partitioners like Spinner~\cite{spinner}, ParMETIS~\cite{10.1145/369028.369103}, XtraPulp~\cite{xtrapulp}, and Sheep~\cite{Margo:2015:SDG:2824032.2824046} are outperformed by DNE in terms of replication factor on a large number of real-world graphs~\cite{dne}.

\emph{(3) Hybrid Partitioning. } We also compare ourselves to Hybrid Edge Partitioner (HEP)~\cite{hep} which operates between in-memory and streaming partitioning using a tunable parameter $\tau$. We set $\tau$ to different values (100, 10, 1) as in the original publication~\cite{hep}. With a high setting of $\tau = 100$, HEP behaves more like an in-memory partitioner, while with a low setting of $\tau = 1$, HEP behaves more like a streaming partitioner. 

\begin{figure*}
	\centering	
	\captionsetup[subfloat]{captionskip=-2pt}
	%\subfloat[CAPTION]{BILDERCODE}\qquad
		\subfloat[OK: Replication factor.]{\includegraphics[width=0.32\textwidth]{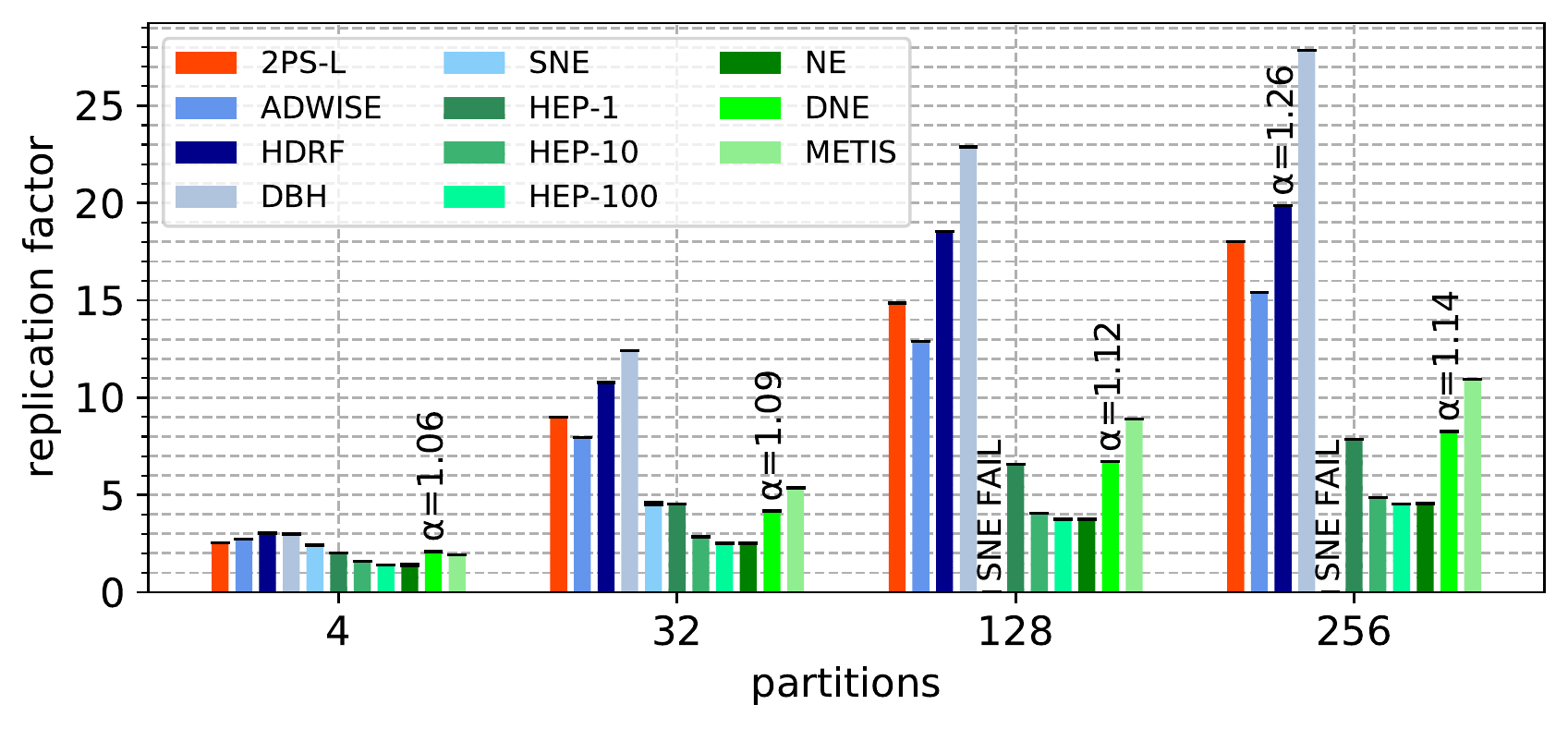}}
	\subfloat[OK: Run-time (logscale).]{\label{b}   \includegraphics[width=0.32\textwidth]{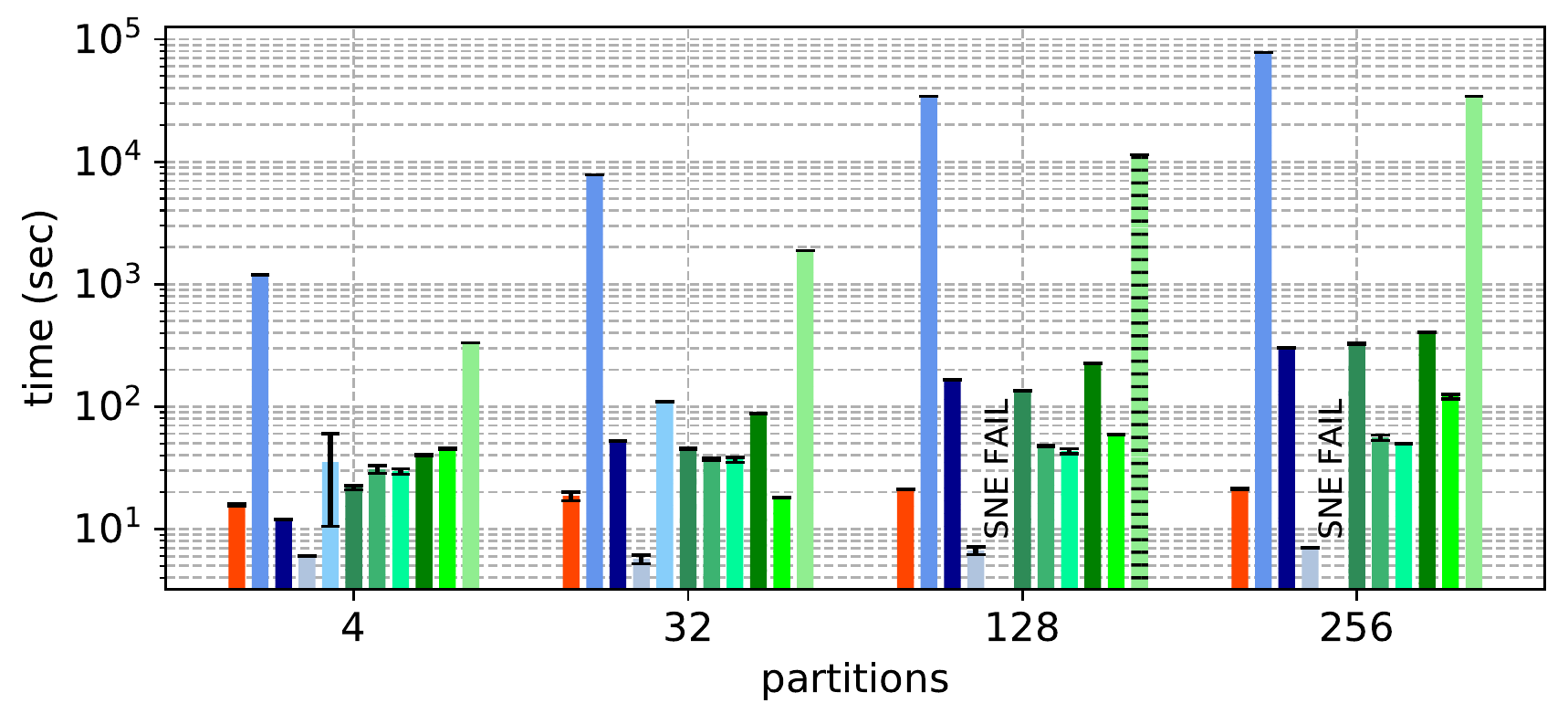}} 
	\subfloat[OK: Memory overhead (logscale).]{\label{c}   \includegraphics[width=0.32\textwidth]{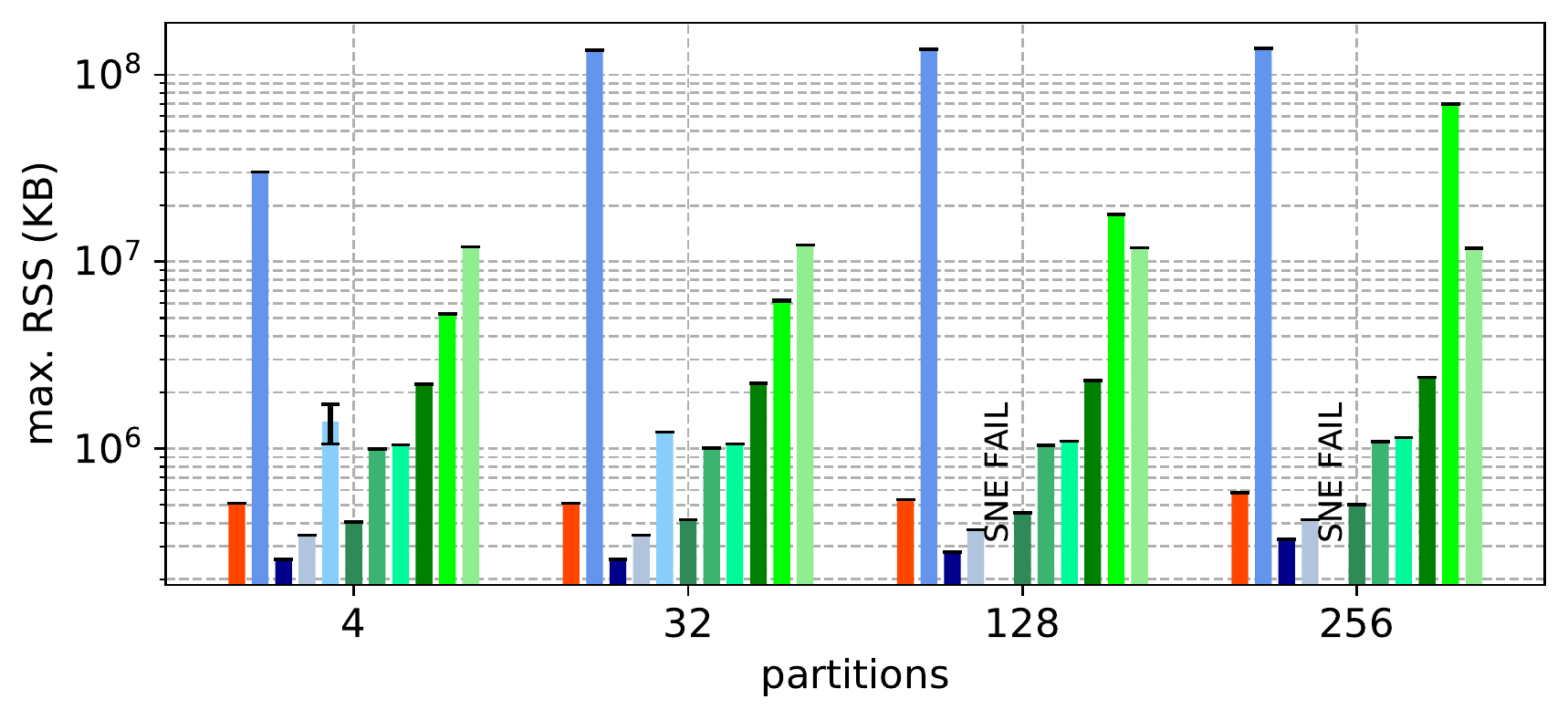}}\\
	\vspace{-0.4cm}
	\subfloat[IT: Replication factor.]{\includegraphics[width=0.32\textwidth]{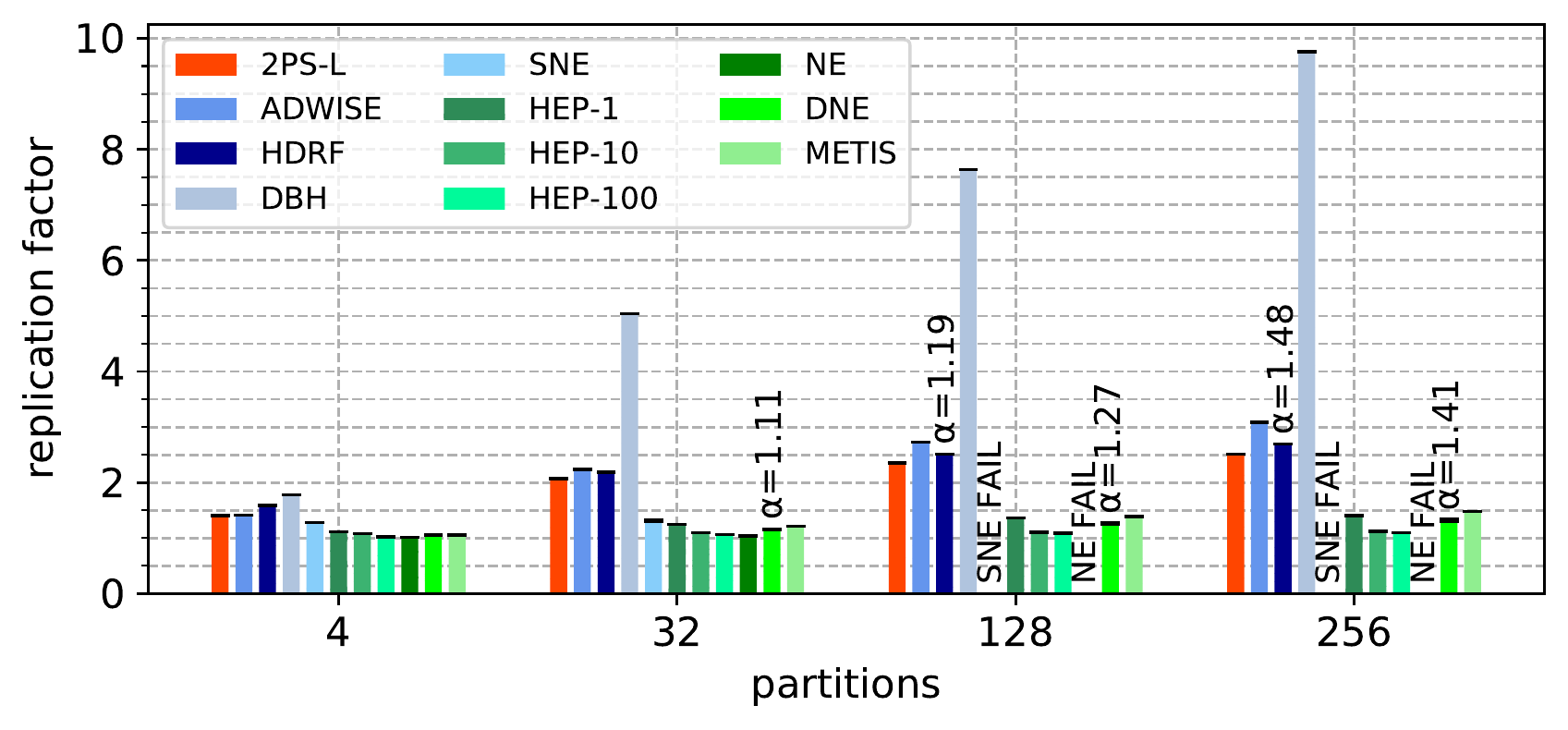}}
	\subfloat[IT: Run-time (logscale).]{\label{b}   \includegraphics[width=0.32\textwidth]{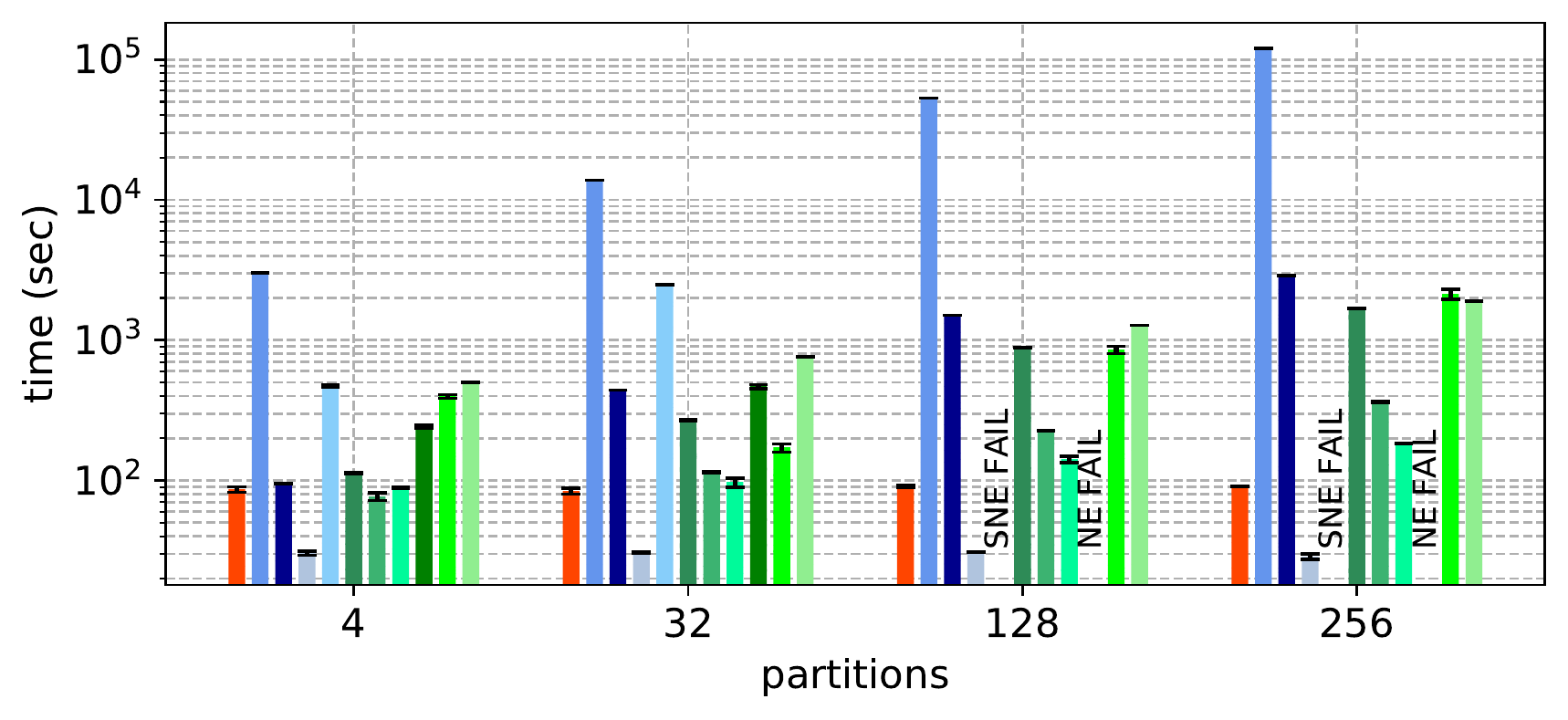}} 
	\subfloat[IT: Memory overhead (logscale).]{\label{c}   \includegraphics[width=0.32\textwidth]{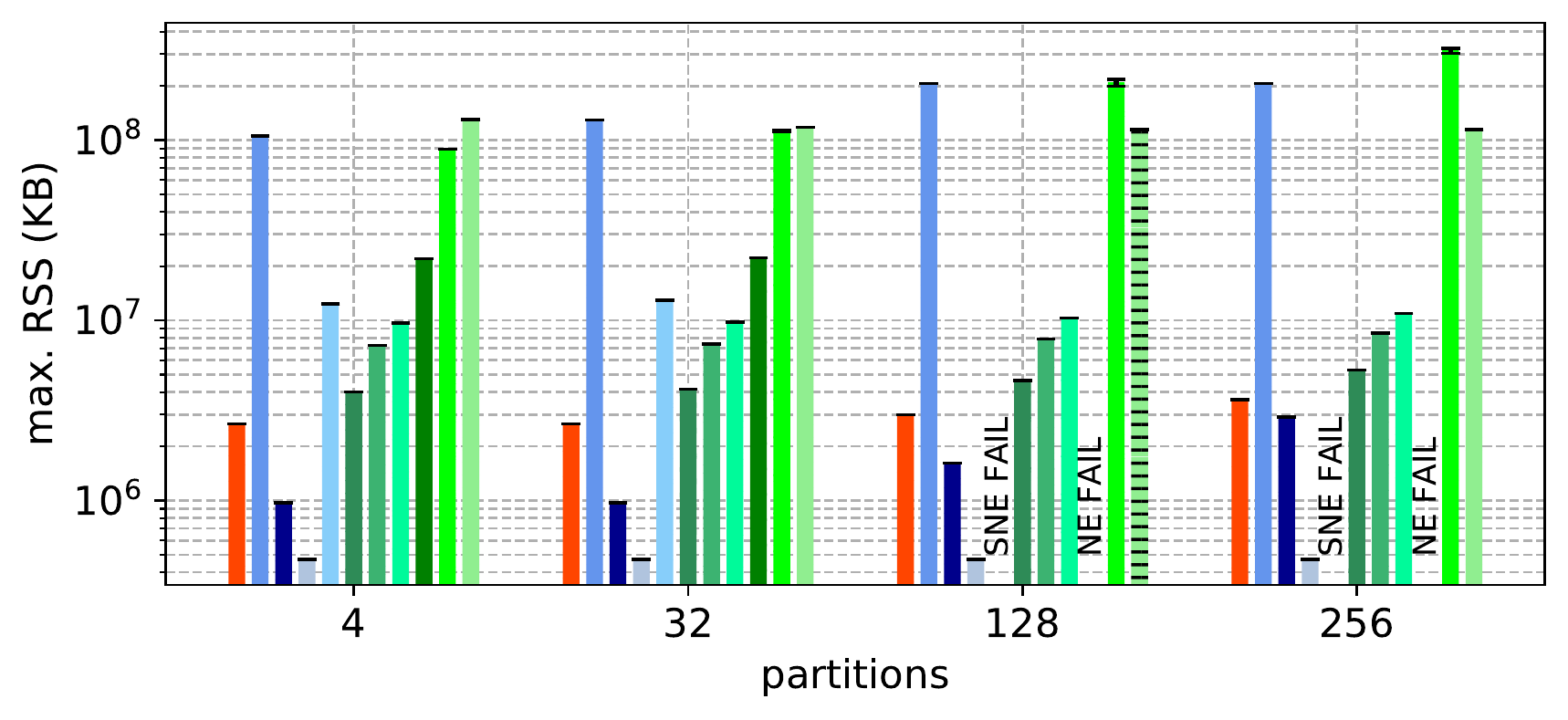}}\\
	\vspace{-0.4cm}
	\subfloat[TW: Replication factor.]{\label{a}   \includegraphics[width=0.32\textwidth]{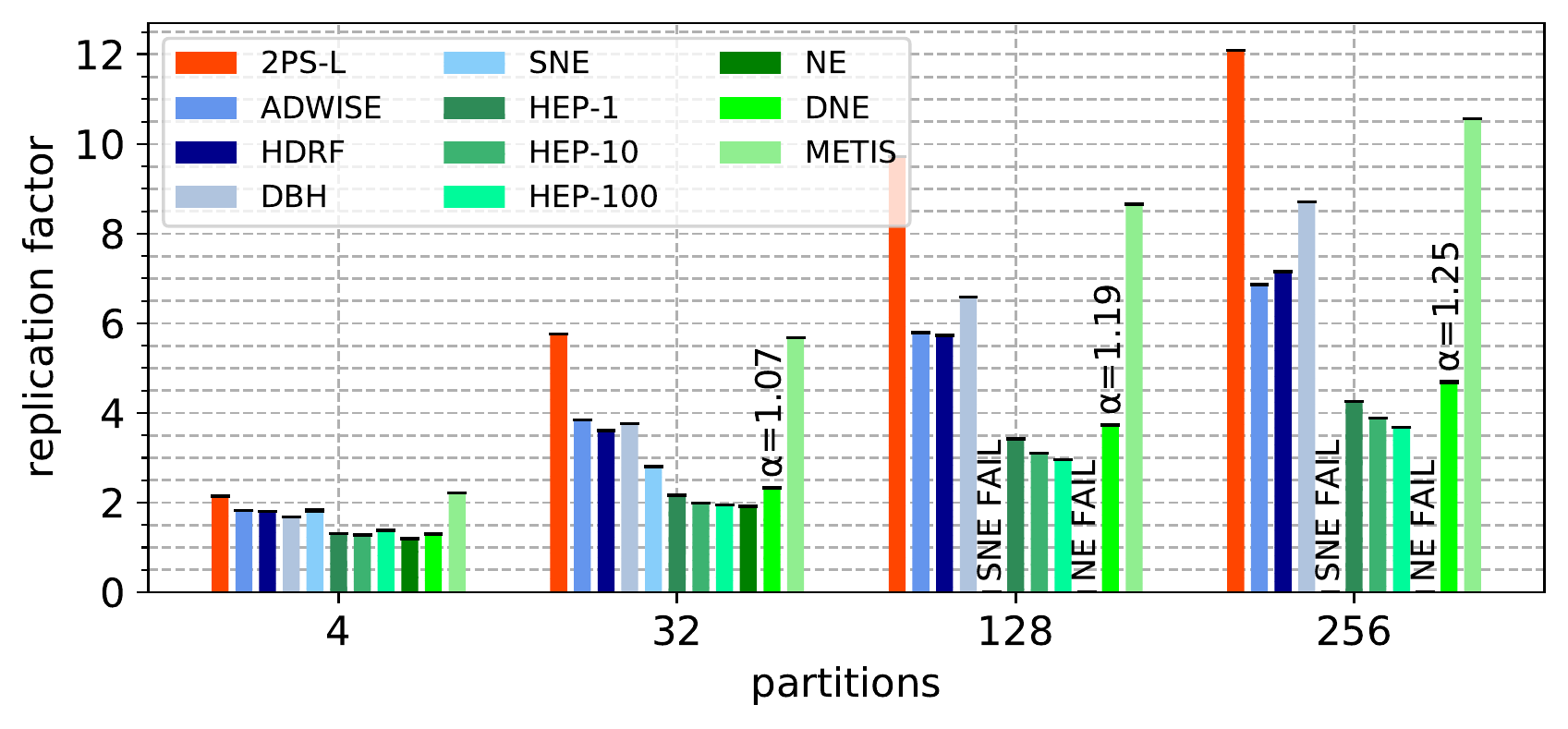}}
	\subfloat[TW: Run-time (logscale).]{\label{b}   \includegraphics[width=0.32\textwidth]{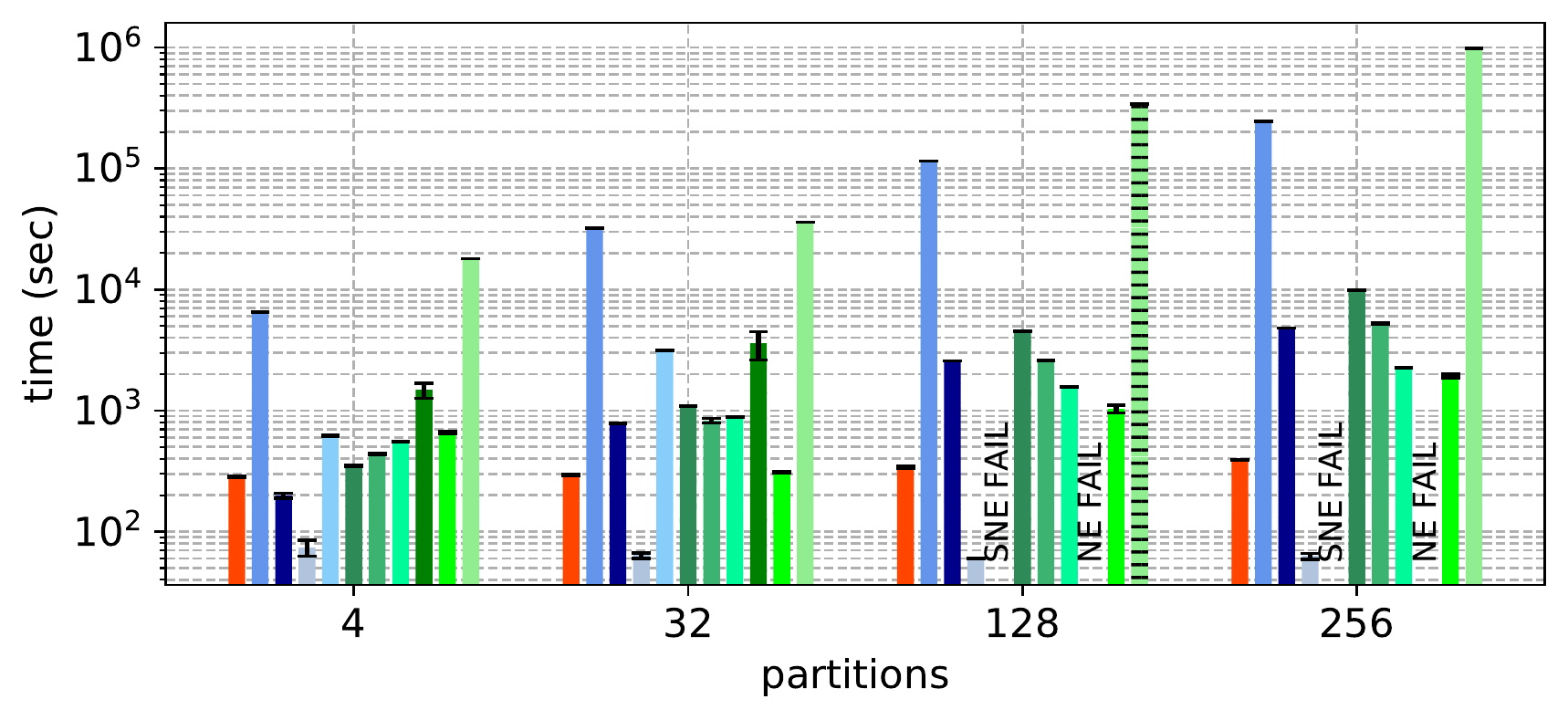}}
	\subfloat[TW: Memory overhead (logscale).]{\label{c}   \includegraphics[width=0.32\textwidth]{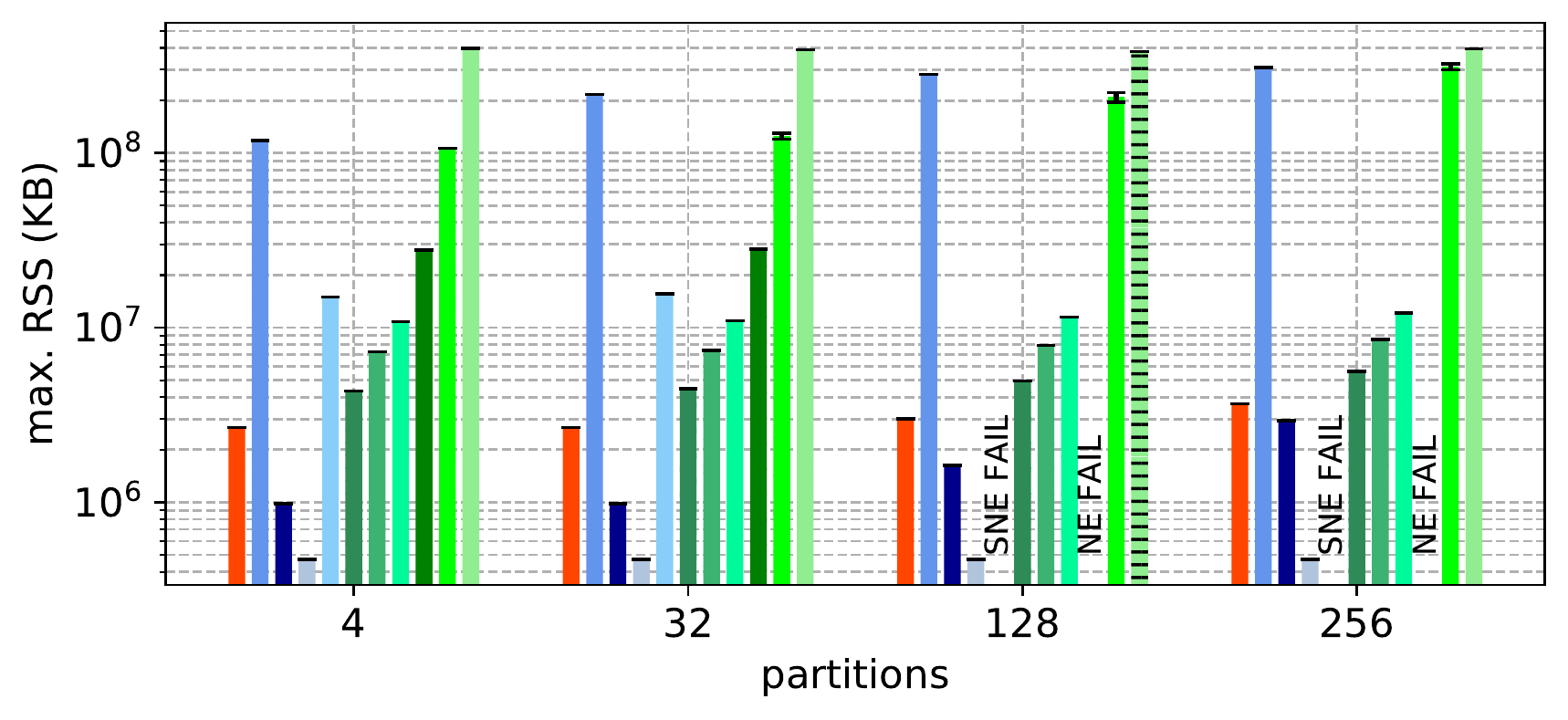}} \\
	\vspace{-0.4cm}
	\subfloat[FR: Replication factor.]{\label{a}   \includegraphics[width=0.32\textwidth]{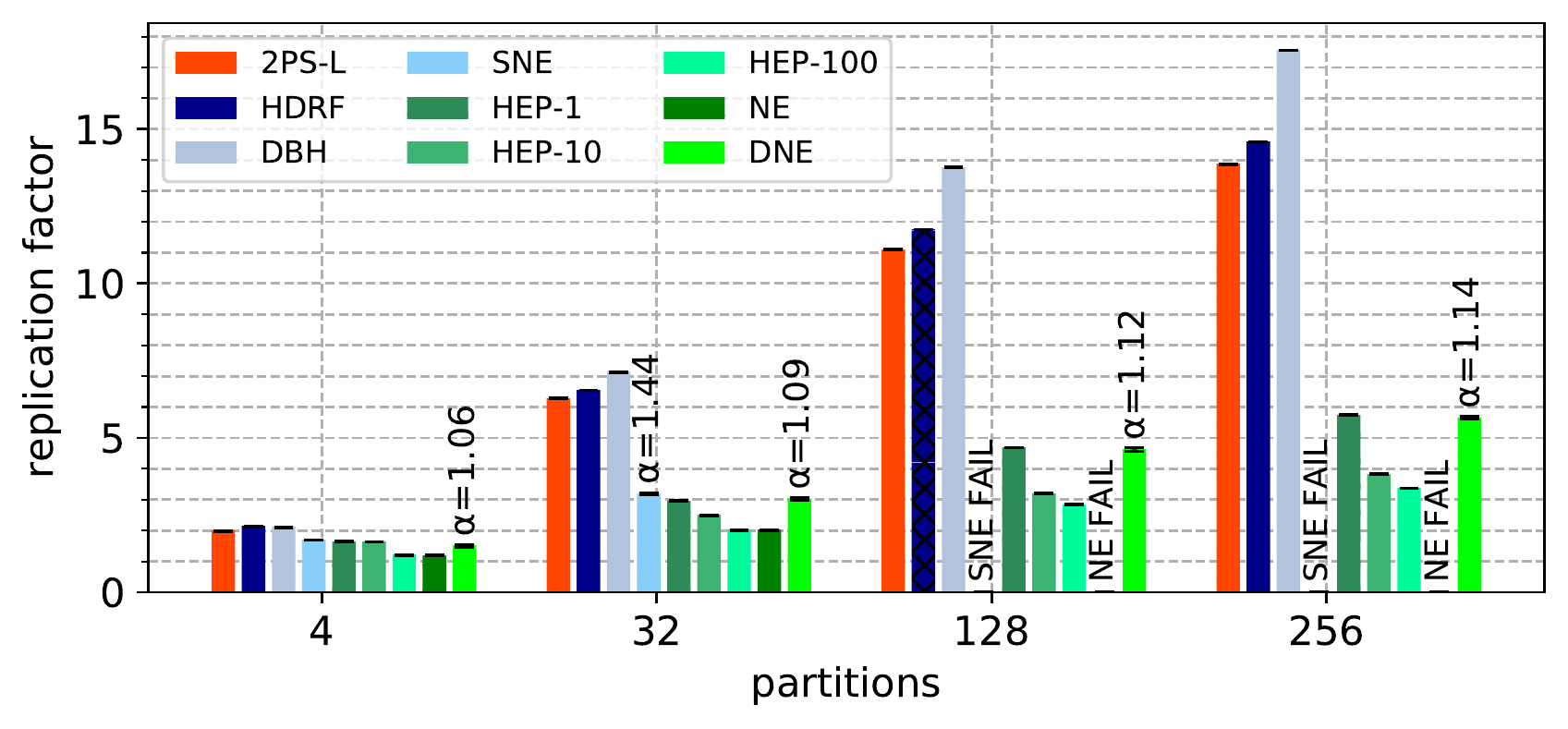}}
	\subfloat[FR: Run-time (logscale).]{\label{b}   \includegraphics[width=0.32\textwidth]{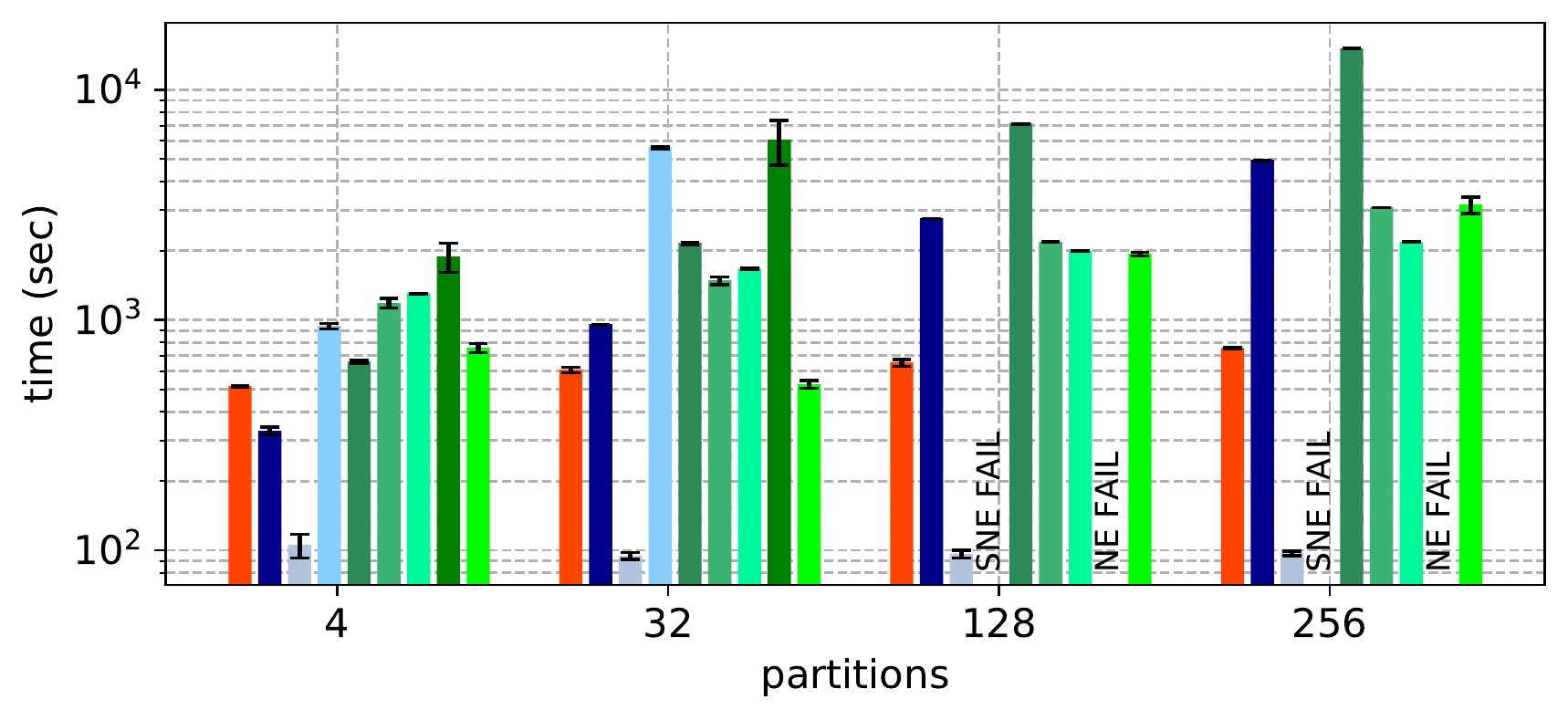}}
	\subfloat[FR: Memory overhead (logscale).]{\label{c}   \includegraphics[width=0.32\textwidth]{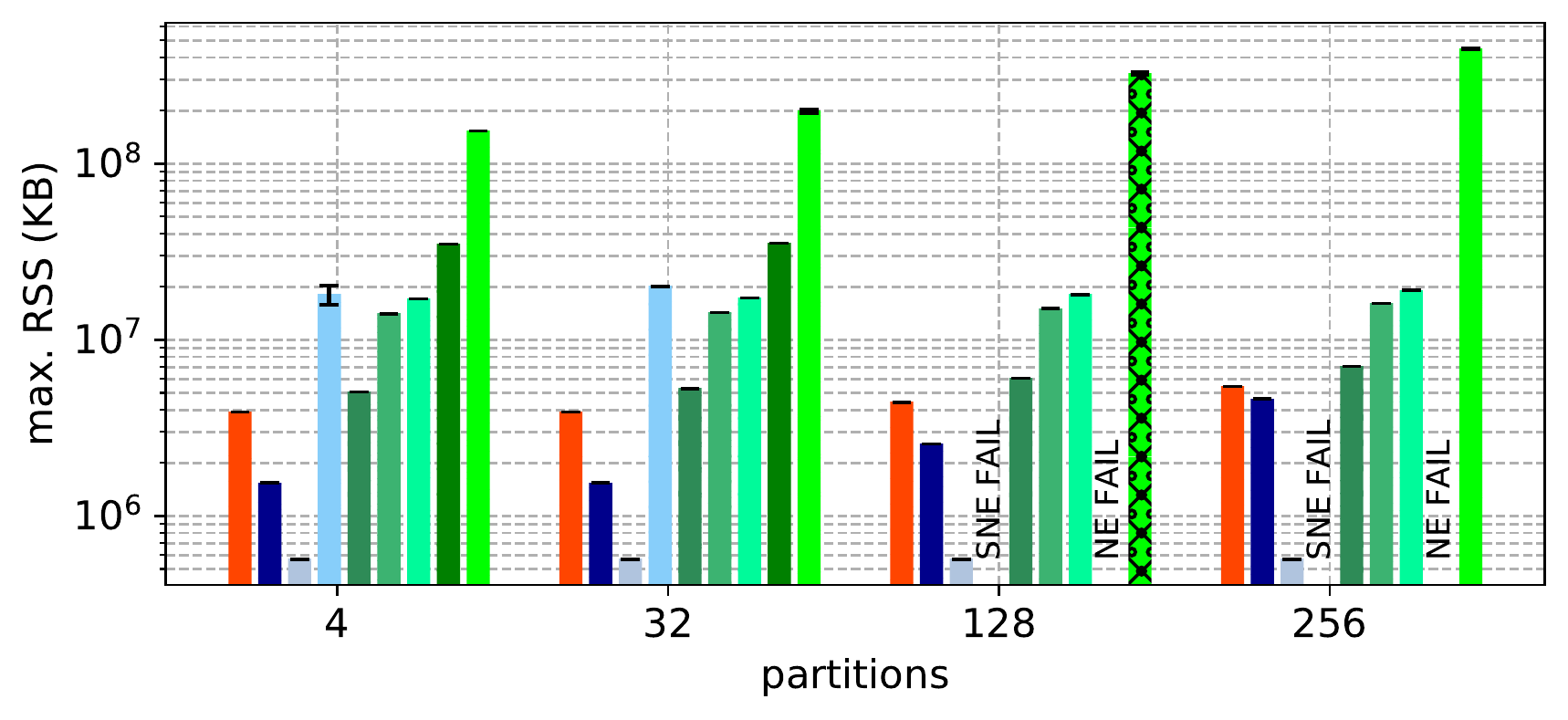}} \\
	\vspace{-0.4cm}
		\subfloat[UK: Replication factor.]{\label{a}   \includegraphics[width=0.32\textwidth]{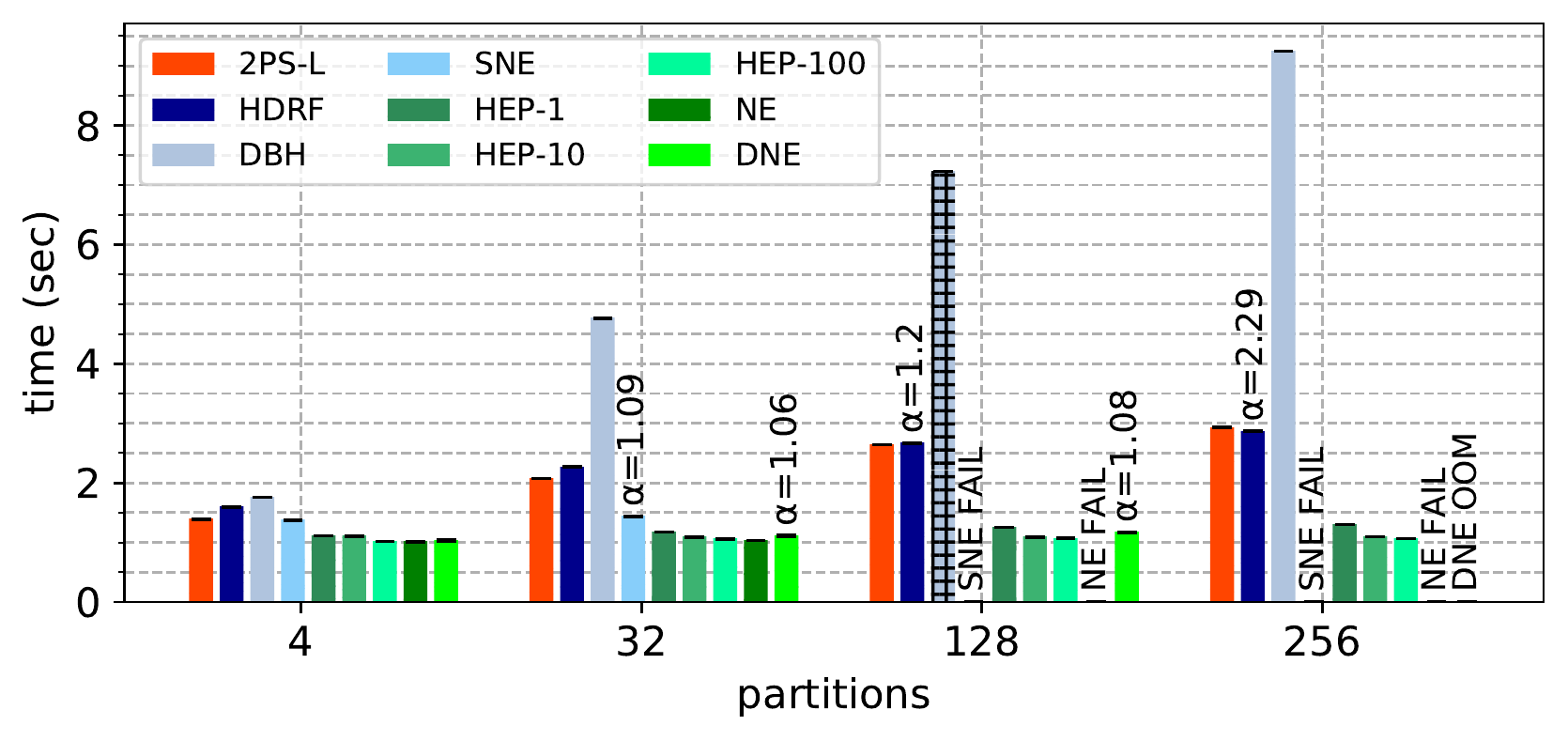}}
	\subfloat[UK: Run-time (logscale).]{\label{b}   \includegraphics[width=0.32\textwidth]{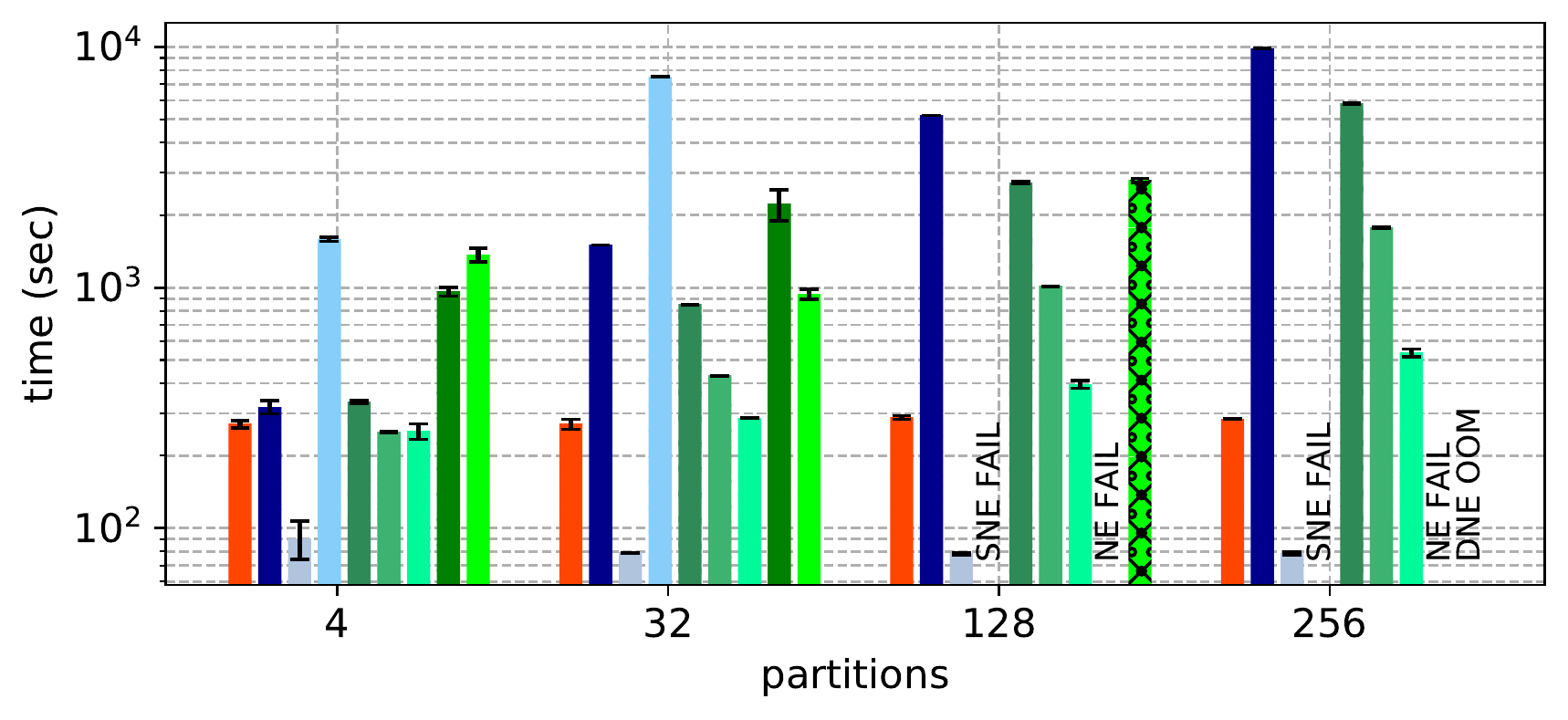}}
	\subfloat[UK: Memory overhead (logscale).]{\label{c}   \includegraphics[width=0.32\textwidth]{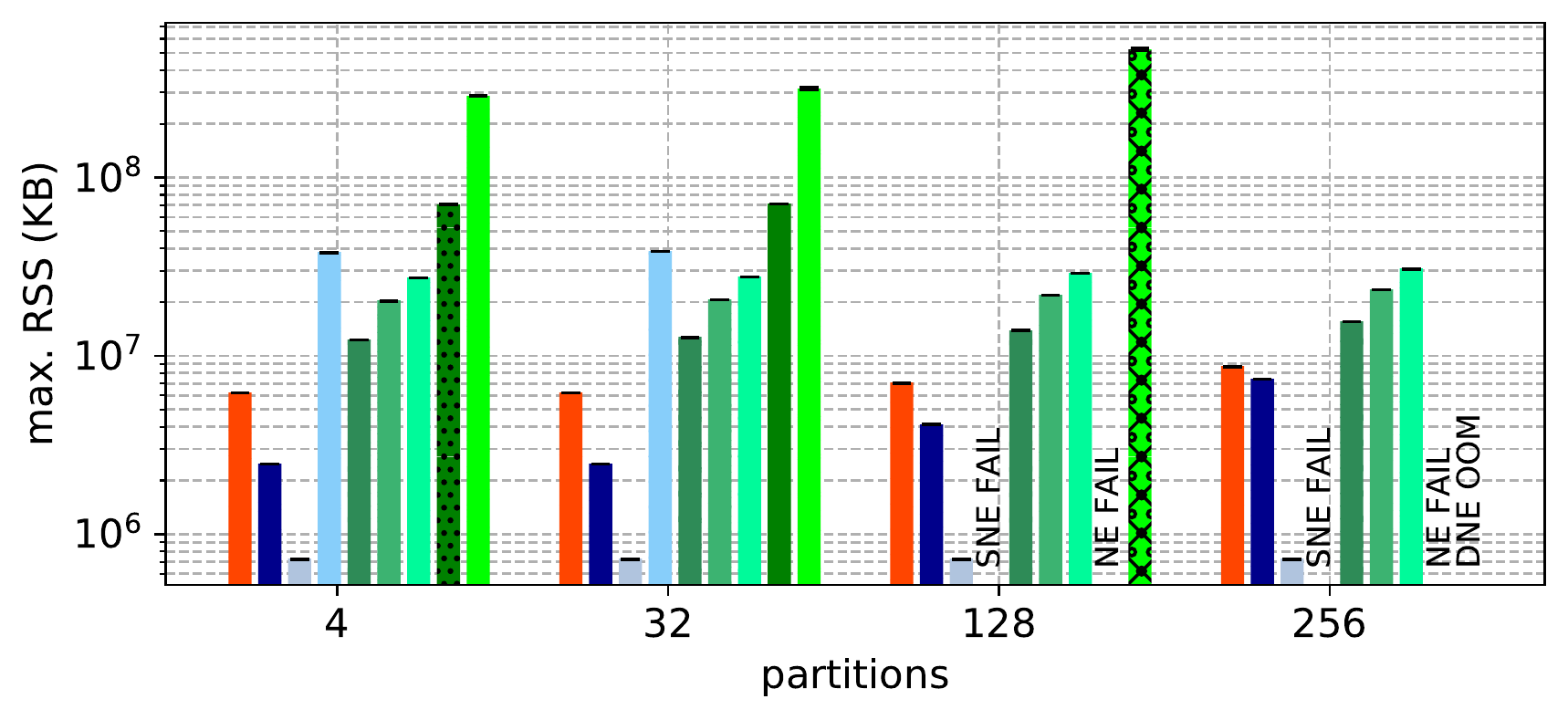}} \\
	\vspace{-0.4cm}
		\subfloat[GSH: Replication factor.]{\label{a}   \includegraphics[width=0.32\textwidth]{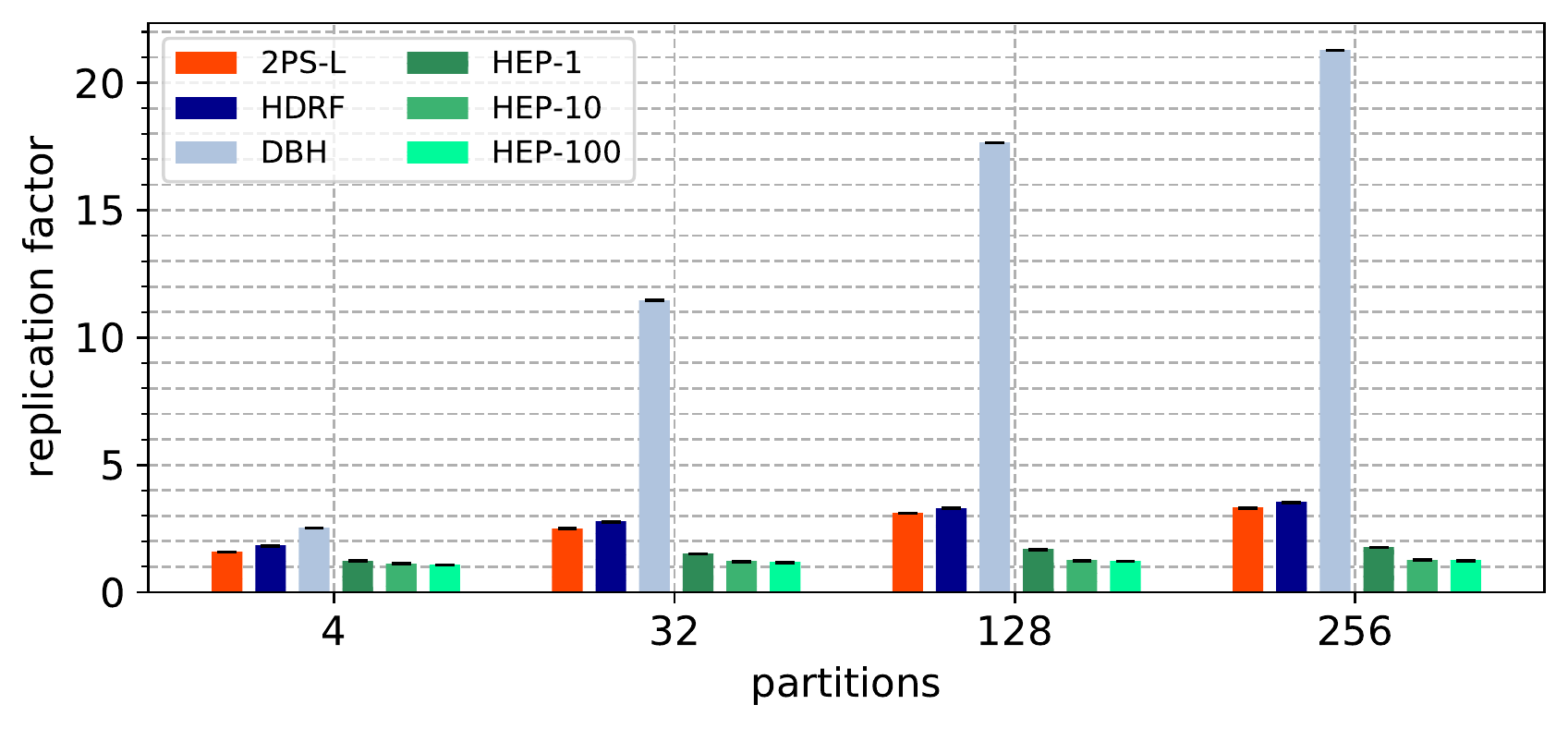}}
	\subfloat[GSH: Run-time (logscale).]{\label{b}   \includegraphics[width=0.32\textwidth]{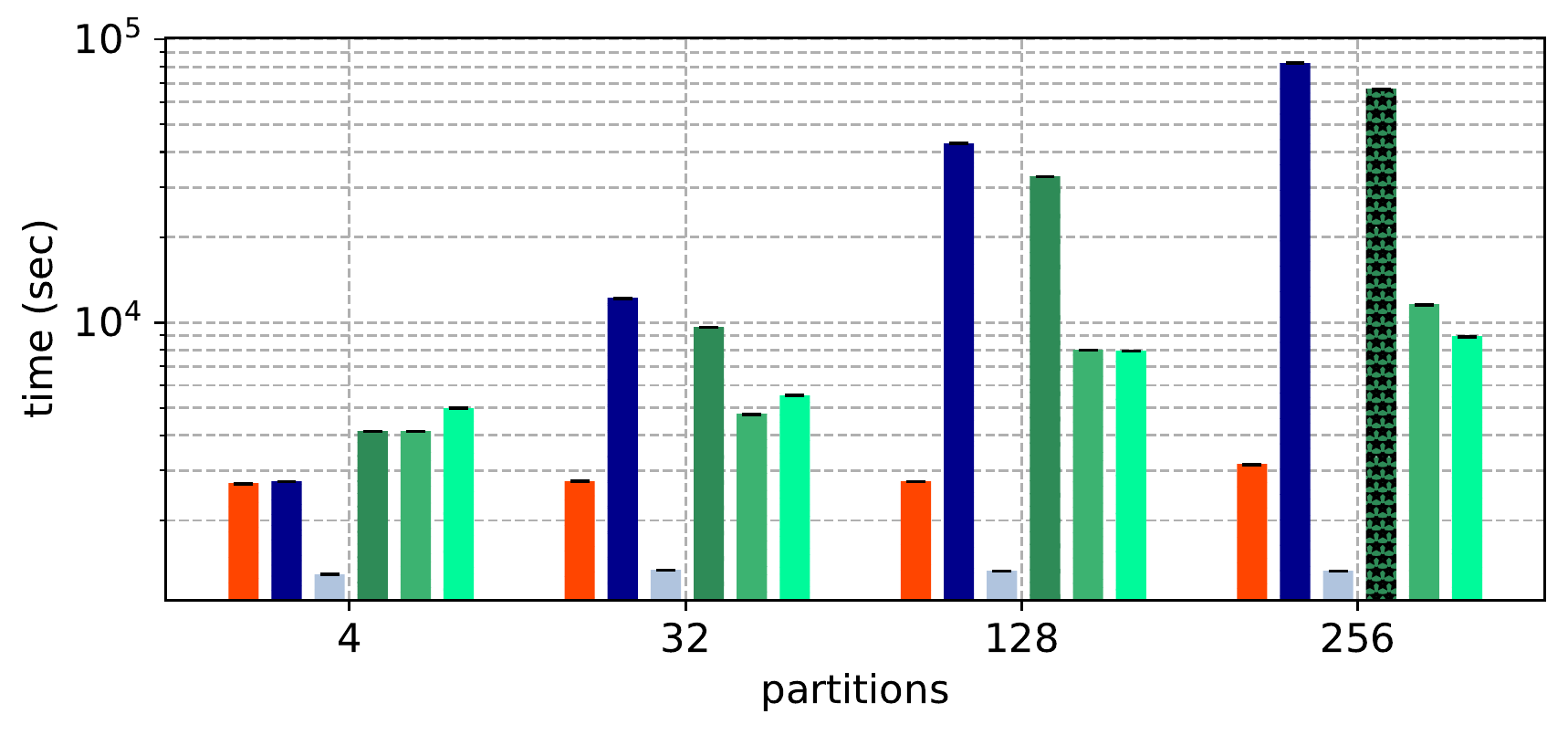}}
	\subfloat[GSH: Mem. overhead (log.).]{\label{c}   \includegraphics[width=0.32\textwidth]{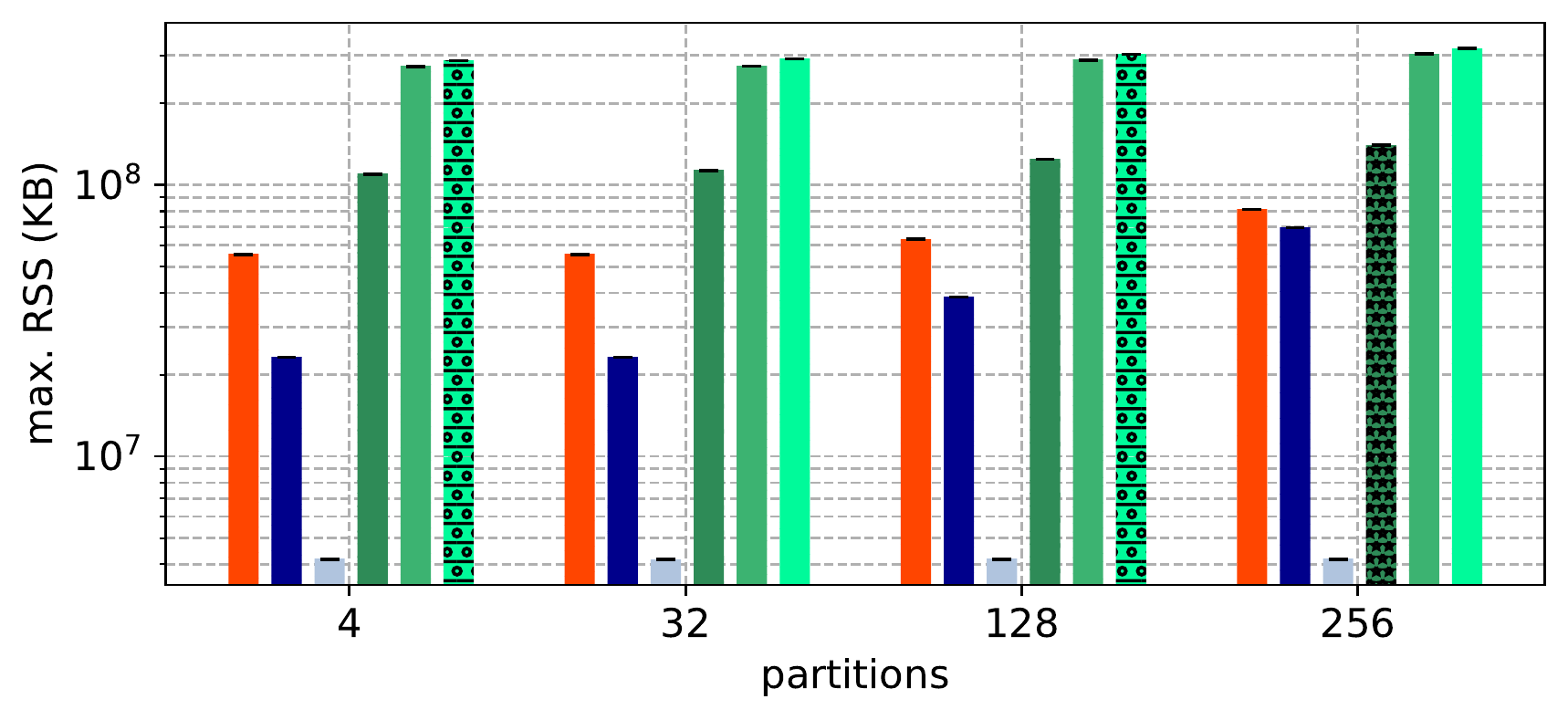}}\\
	\vspace{-0.4cm}
		\subfloat[WDC: Replication factor.]{\label{a}   \includegraphics[width=0.32\textwidth]{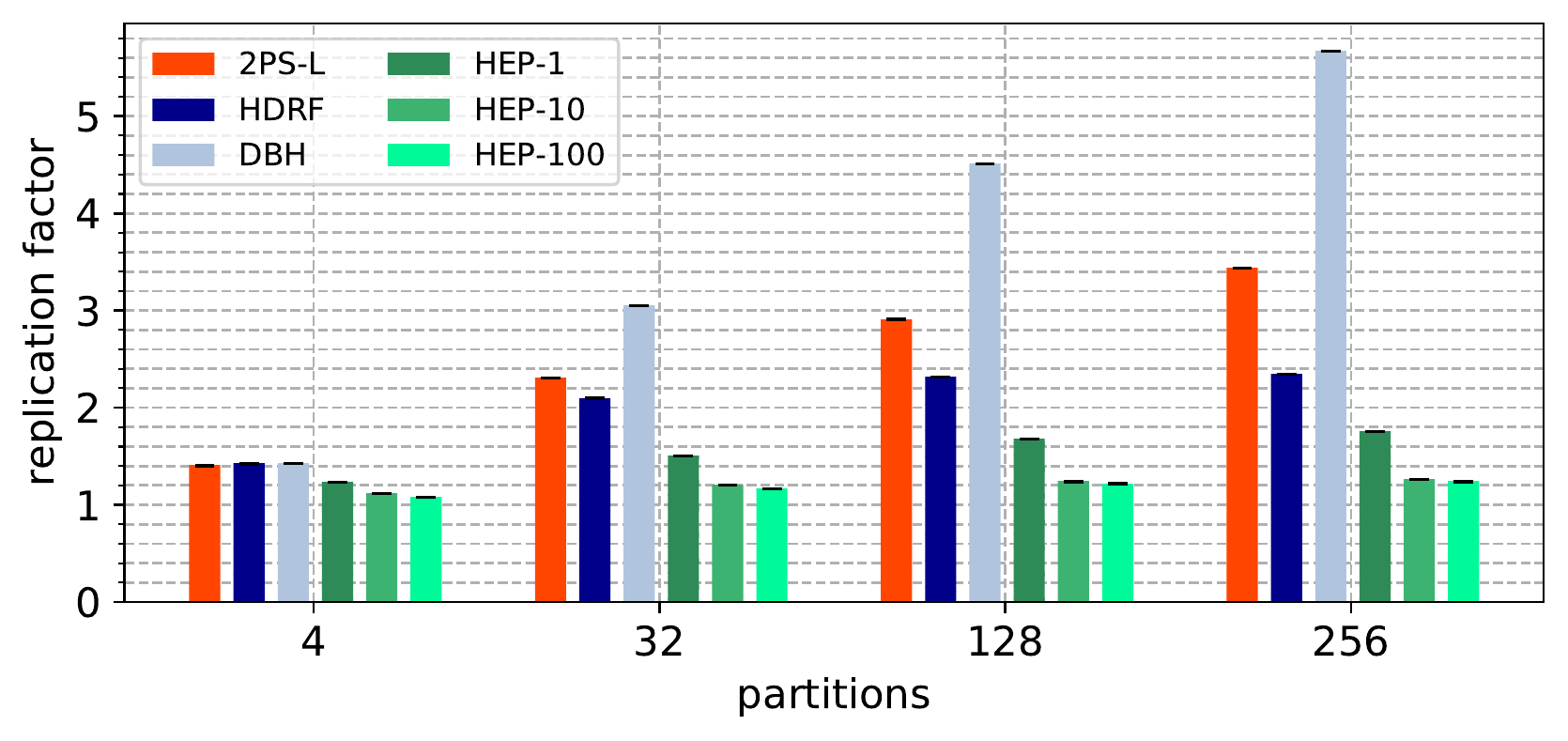}}
	\subfloat[WDC: Run-time (logscale).]{\label{b}   \includegraphics[width=0.32\textwidth]{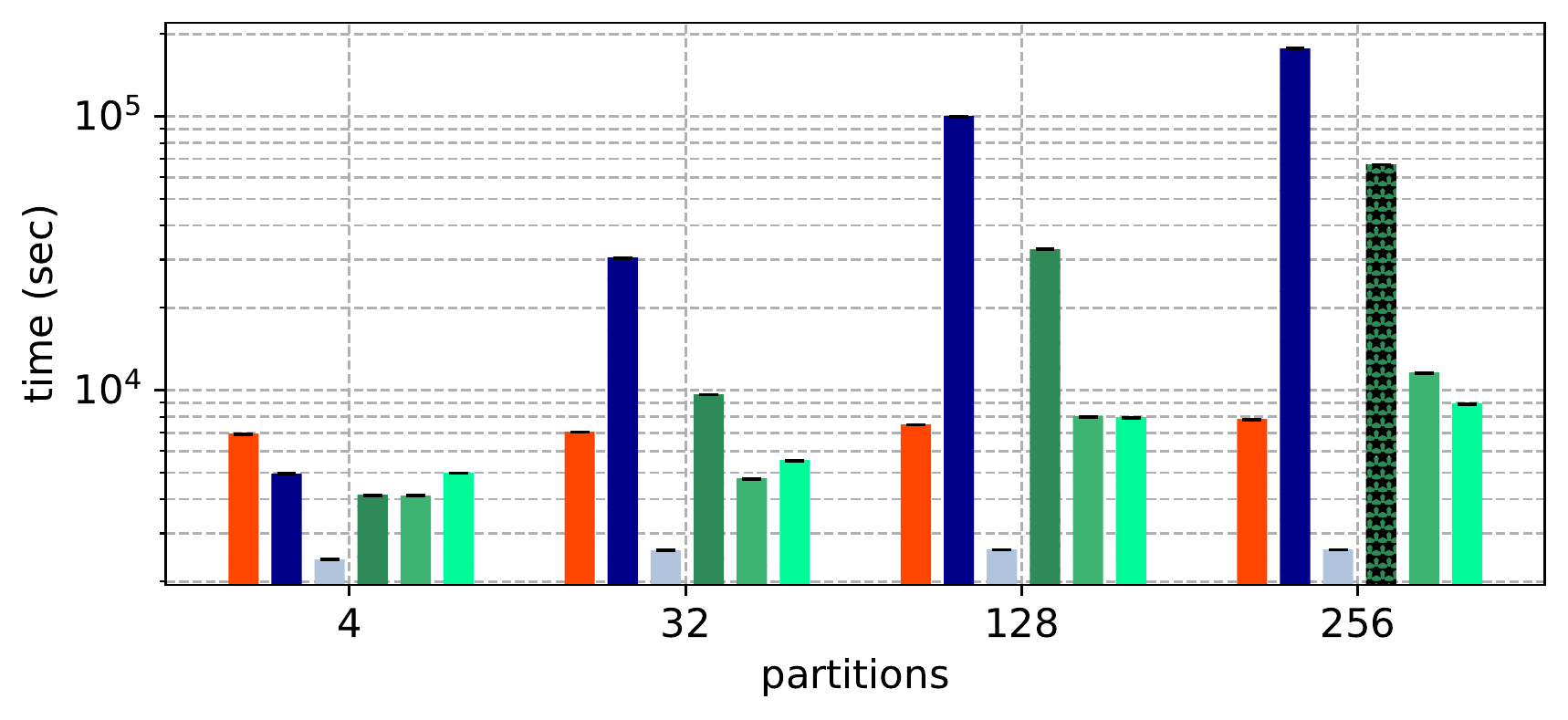}}
	\subfloat[WDC: Mem. overhead (log.).]{\label{c}   \includegraphics[width=0.32\textwidth]{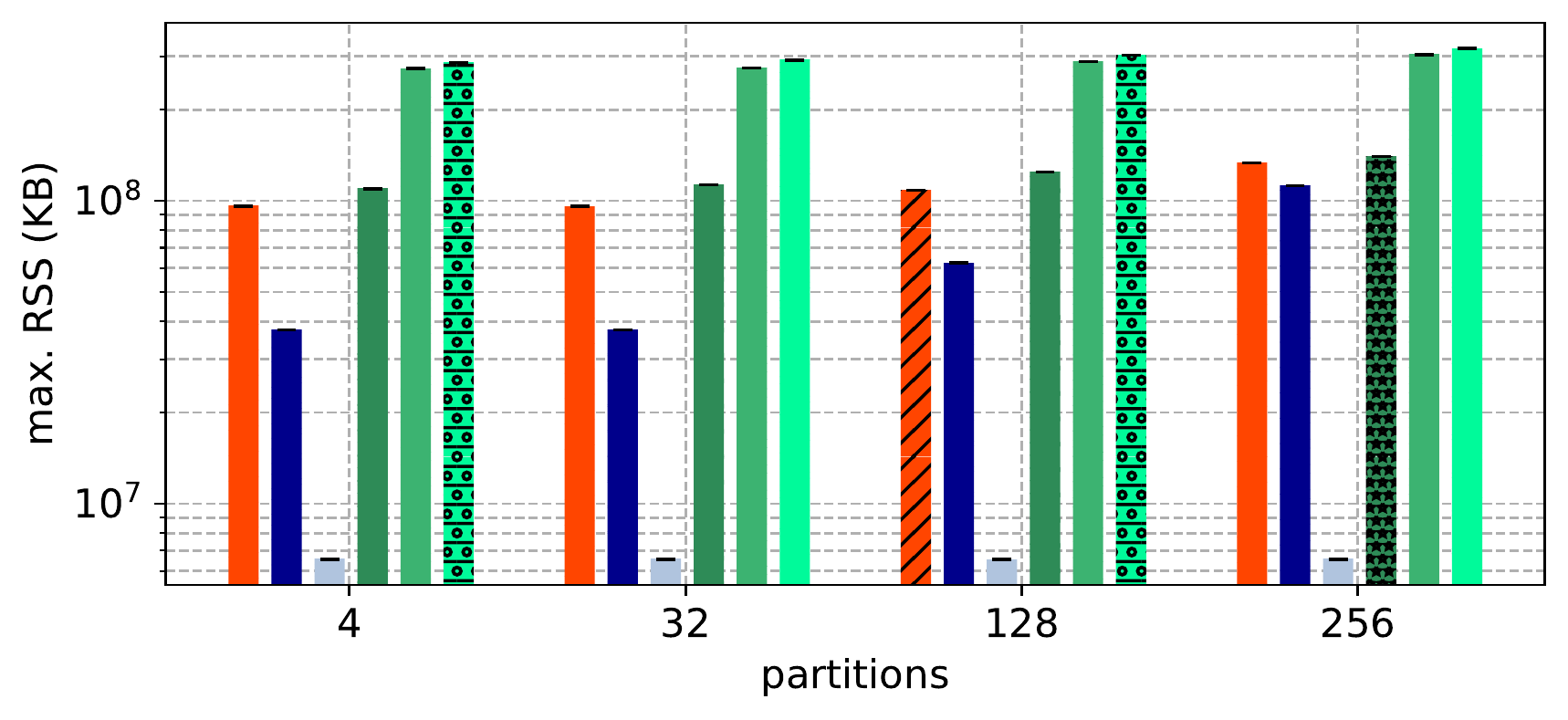}}
	%\vspace{-5pt}
	\caption{Performance results on real-world graphs. }
	\label{eval:perf}
	\vspace{-10pt}
\end{figure*}

\begin{figure*}
\centering
\begin{minipage}[t]{.23\textwidth}
  \centering
  \includegraphics[width=\linewidth]{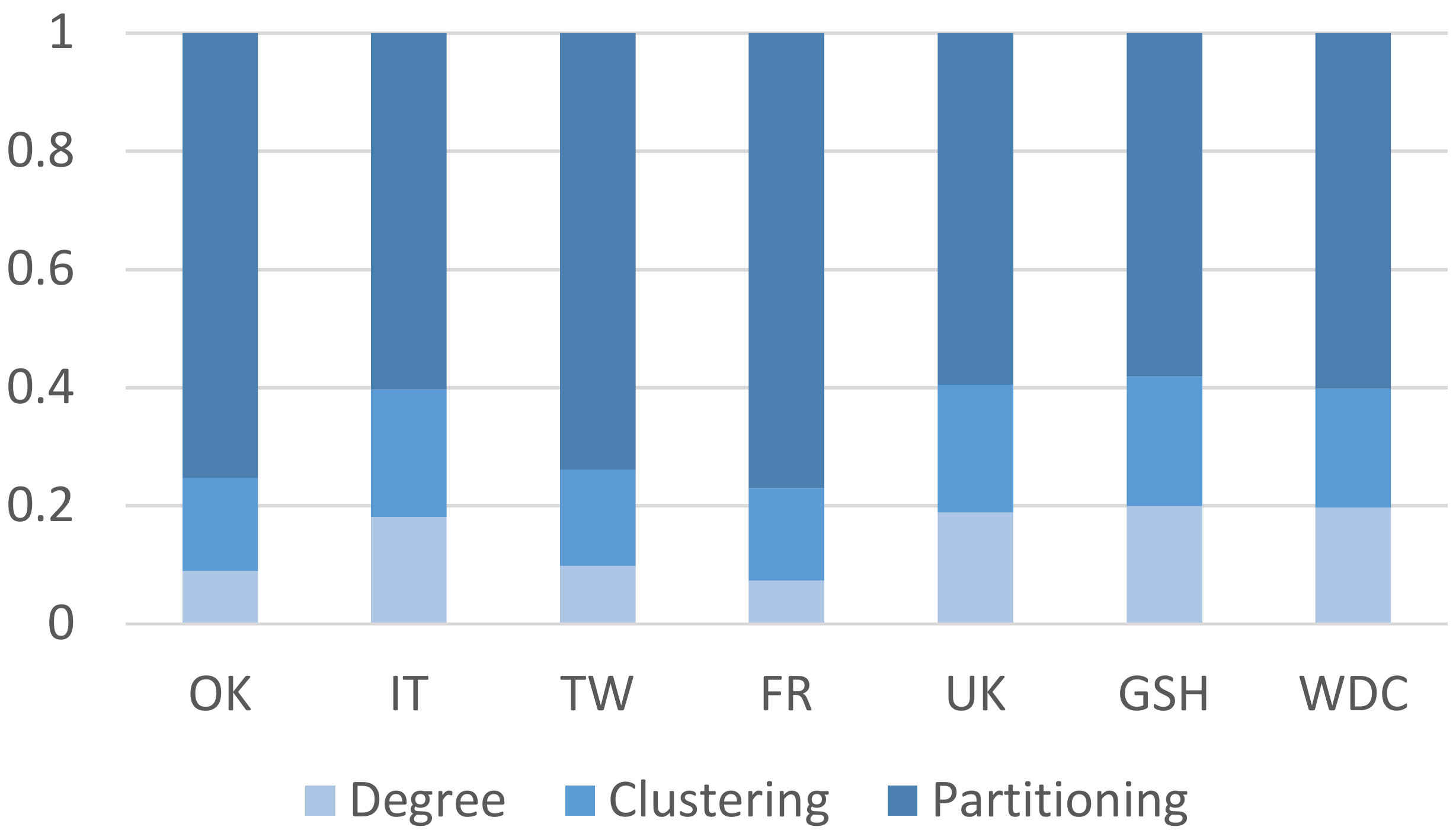}
  \caption{Relative run-time of different phases of 2PS-L (w/o re-streaming) at $k=32$.}
  \label{eval:phases}
\end{minipage}\quad
\begin{minipage}[t]{.23\textwidth}
  \centering
  \includegraphics[width=\linewidth]{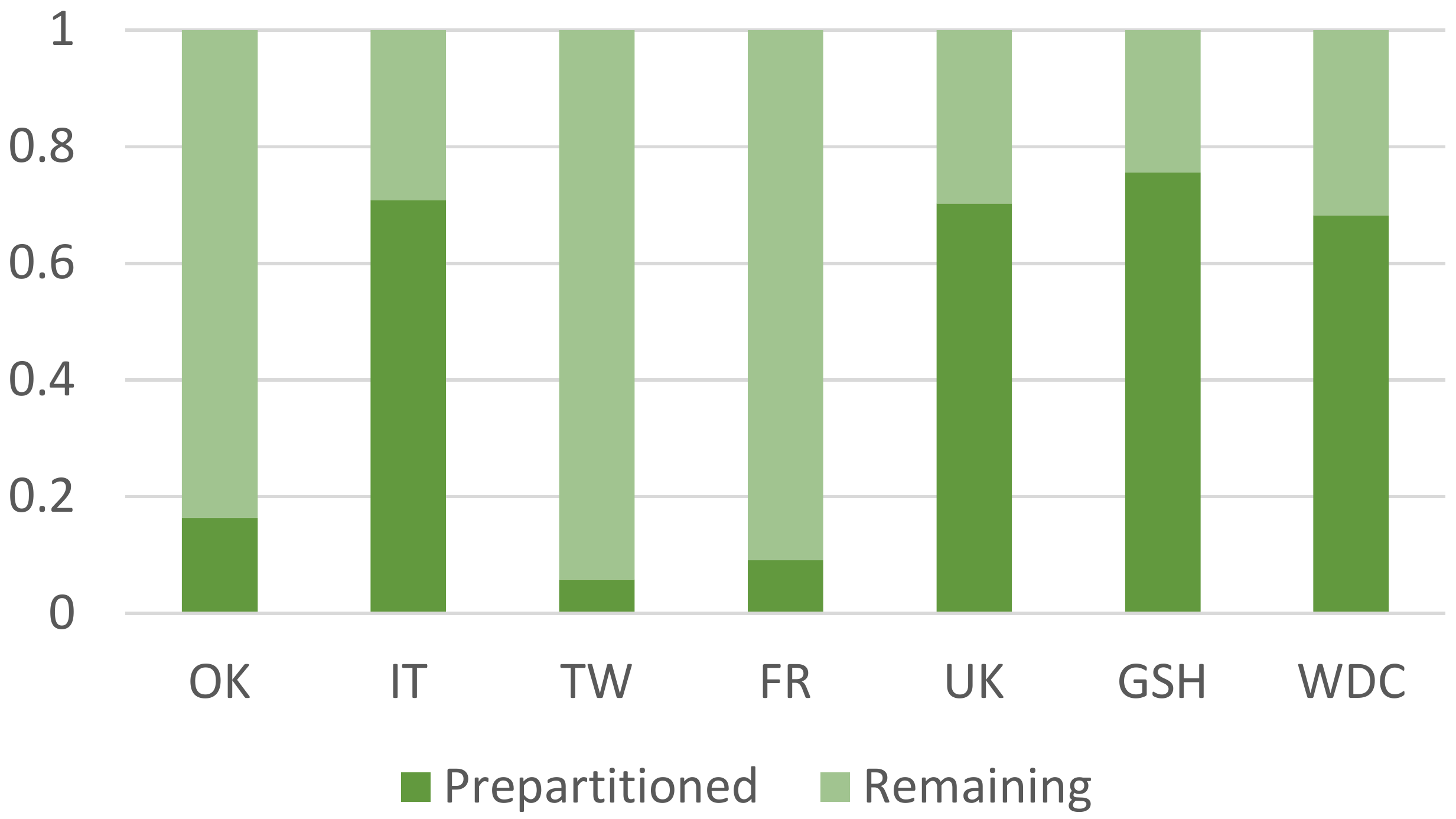}
  \caption{Ratio of prepartitioned vs. remaining edges in 2PS-L (w/o re-streaming) at $k=32$.}
  \label{eval:edge_ratio}
\end{minipage}\hfill\quad 
\begin{minipage}[t]{.23\textwidth}
  \centering
  \includegraphics[width=\linewidth]{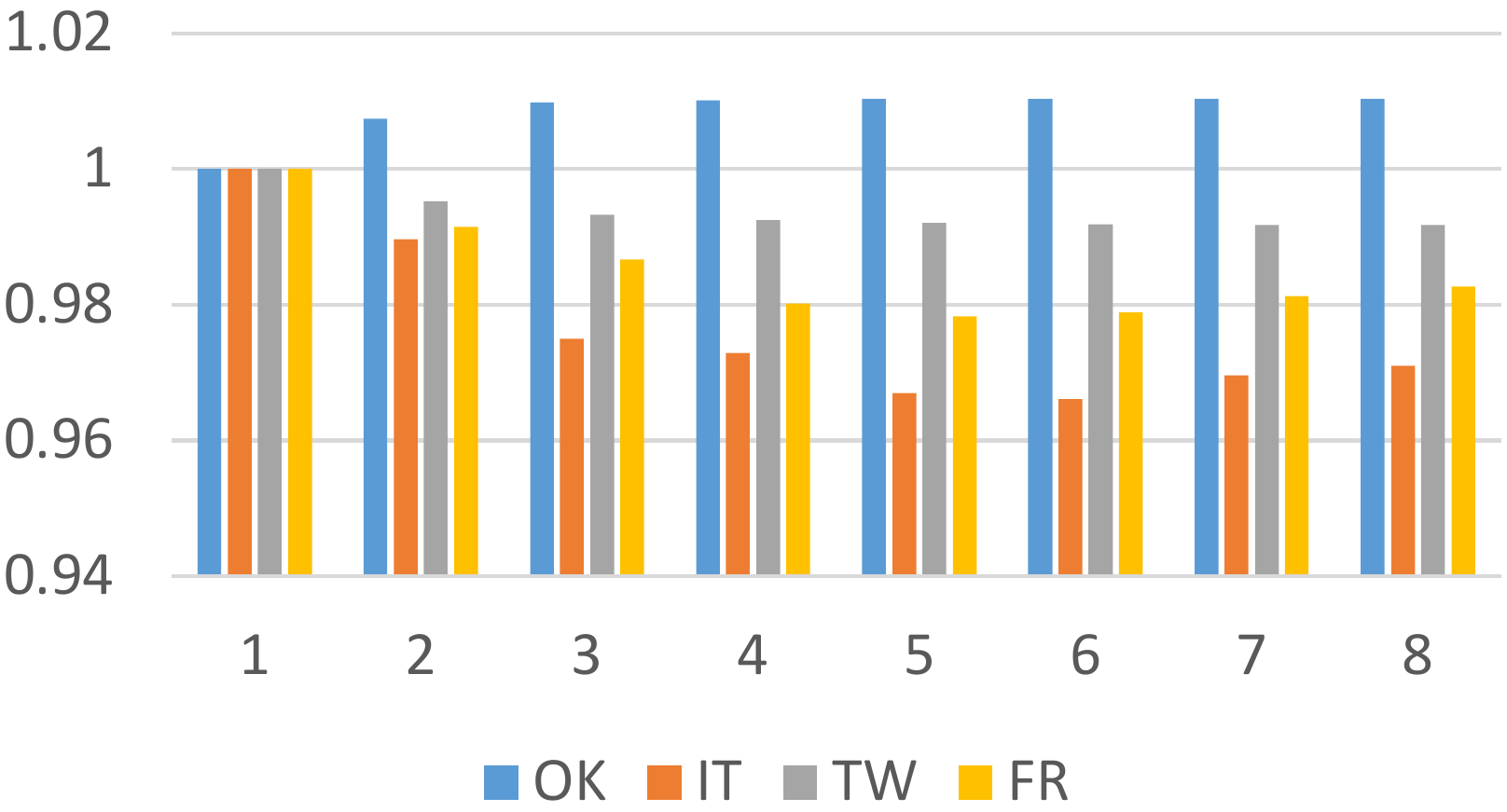}
  \caption{Normalized replication factor (y-axis) vs. streaming clustering passes (x-axis) at $k=32$.}
  \label{eval:restreaming_rf}
\end{minipage}\quad
\begin{minipage}[t]{.23\textwidth}
  \centering
  \includegraphics[width=\linewidth]{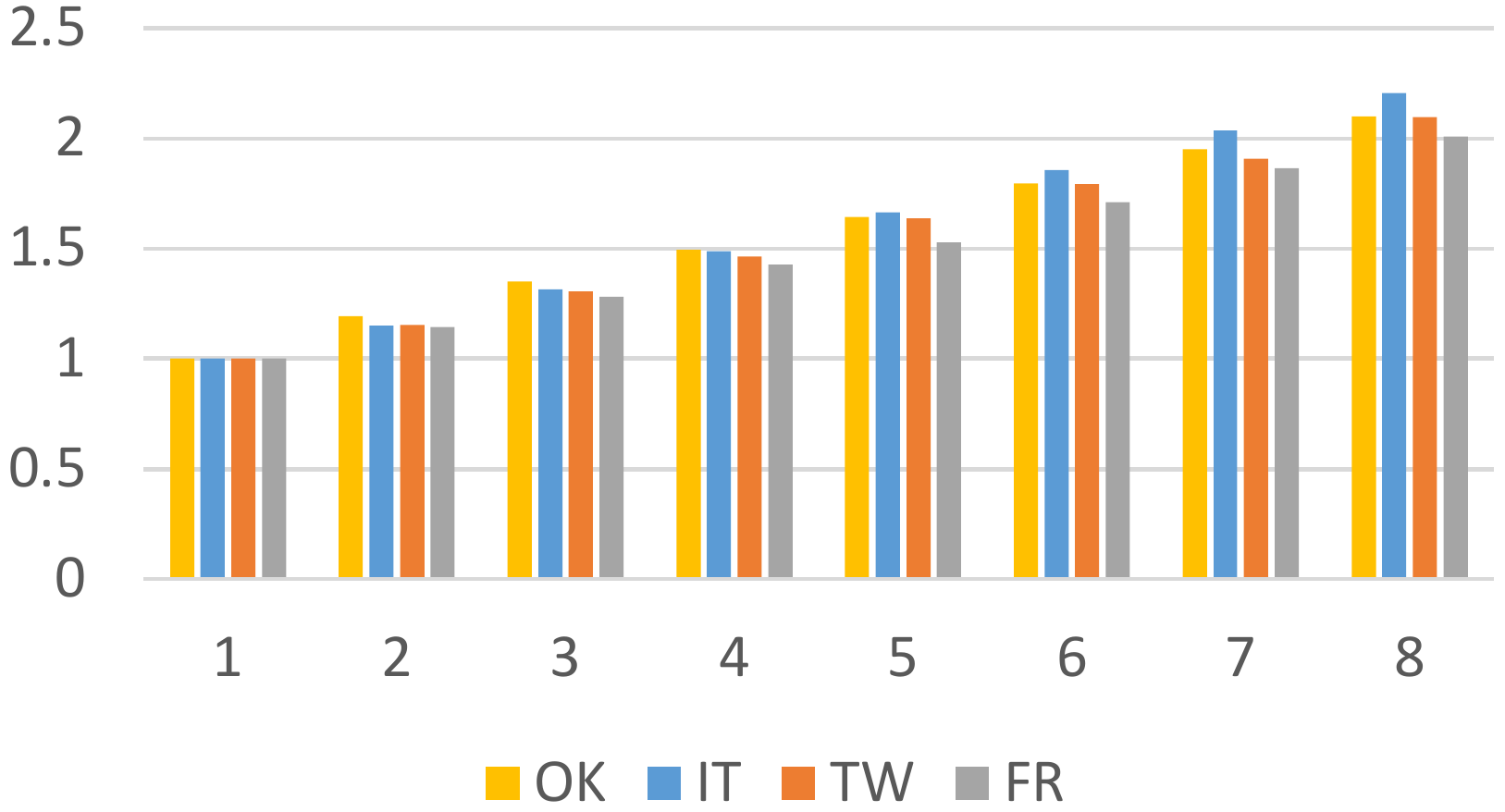}
  \caption{Normalized total 2PS-L run-time (y-axis) vs. streaming clustering passes (x-axis) at $k=32$.}
  \label{eval:restreaming_runtime}
\end{minipage}
\vspace{-10pt}
\end{figure*}

We discuss other partitioners extensively in Section~\ref{sec:related} and detail our experimental settings in the appendix.

\subsection{Partitioning of Real-World Graphs}
\label{sec:realworld}

We perform our experiments for $k = \{4, 32, 128, 256\}$ partitions. We repeat each experiment 3 times and report the mean value along with error bars that show the standard deviation. The key performance metrics we report are replication factor, partitioning run-time, and memory overhead. We also track balancing. In most cases, the balancing constraint $\alpha = 1.05$ is met by all partitioners; if this is not the case, we report the measured $\alpha$ in the plot. The results comprise \emph{all} costs of 2PS-L, including any preprocessing.

To allow for a separate analysis of the effects of I/O speeds on the performance of 2PS-L, in the following experiments, we perform several subsequent runs, so that the graph data is factually cached by the operating system in memory. We also perform evaluations with disabled caching in Section~\ref{sec:external} to evaluate the effect of I/O bottlenecks in memory-constrained scenarios.

\emph{Main Observations.}
In Figure~\ref{eval:perf}, we depict all performance measurements. Our main observations are as follows.

(1) The run-time of 2PS-L is independent of the number of partitions. Therefore, 2PS-L is significantly faster than all other stateful partitioners (streaming as well as in-memory partitioners) at higher number of partitions ($k=128$ and $k=256$). For instance, at $k=256$ on the TW graph, 2PS-L is $5.8 \times$ faster than HEP-100, $13.4 \times $ faster than HEP-10, $25.7 \times $ faster than HEP-1, $630 \times $ faster than ADWISE, $12.3 \times $ faster than HDRF, $2500 \times $ faster than METIS, and $5.0 \times $ faster than DNE. Even at $k=4$ and $k=32$, 2PS-L is the fastest stateful partitioner in almost all cases. Only DBH, a stateless partitioner based on hashing, is faster than 2PS-L. Hence, 2PS-L is the first stateful partitioner with a run-time that is competitive to stateless partitioning. This way, 2PS-L can be used in scenarios where existing heavy-weight stateful partitioning would not pay off in an end-to-end comparison when considering the sum of partitioning and subsequent distributed graph processing run-time.

\begin{figure*}
	\centering	
	\captionsetup[subfloat]{captionskip=-2pt}
	%\subfloat[CAPTION]{BILDERCODE}\qquad
	\subfloat[OK: Rep. factor.]{\includegraphics[width=0.12\textwidth]{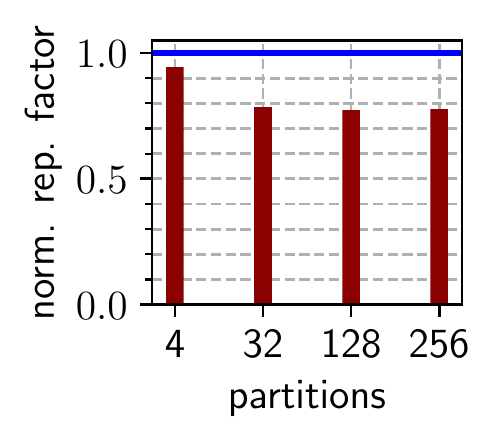}}
	\subfloat[OK: Run-time.]{\label{b}   \includegraphics[width=0.12\textwidth]{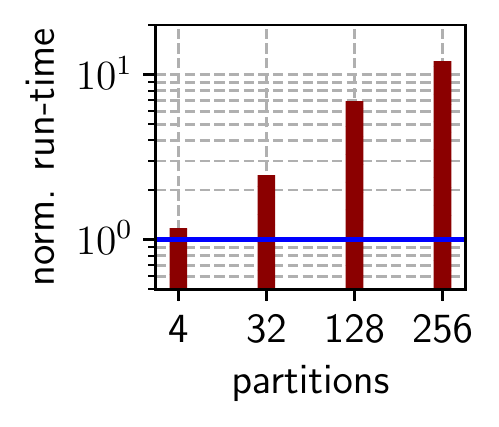}}
	\subfloat[IT: Rep. factor.]{\includegraphics[width=0.12\textwidth]{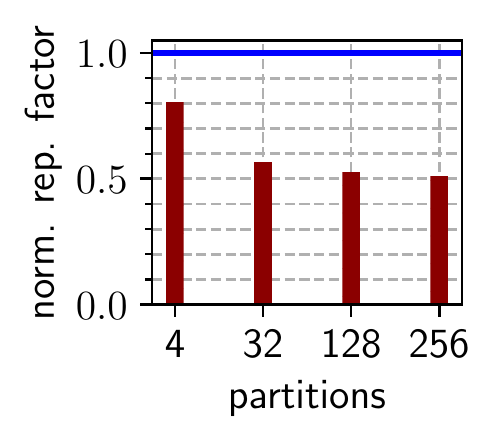}}
	\subfloat[IT: Run-time.]{\label{b}   \includegraphics[width=0.12\textwidth]{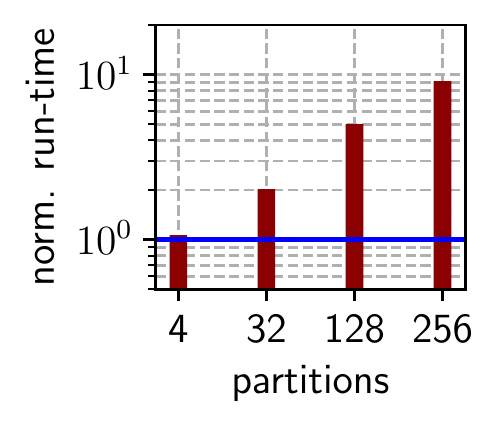}}
	\subfloat[TW: Rep. factor.]{\label{a}   \includegraphics[width=0.12\textwidth]{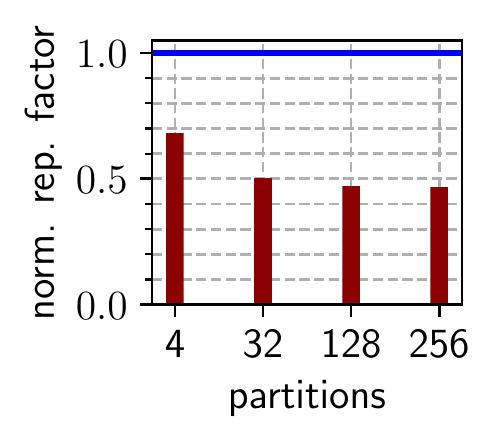}}
	\subfloat[TW: Run-time.]{\label{b}   \includegraphics[width=0.12\textwidth]{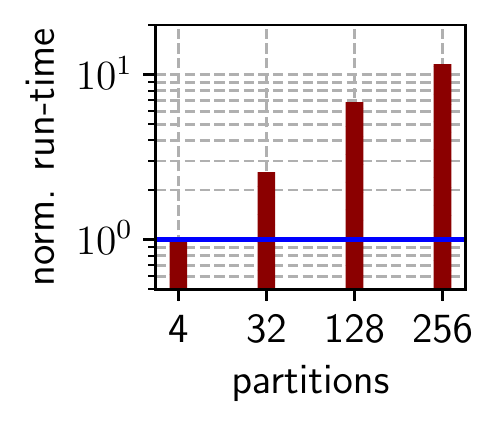}}
	\subfloat[FR: Rep. factor.]{\label{a}   \includegraphics[width=0.12\textwidth]{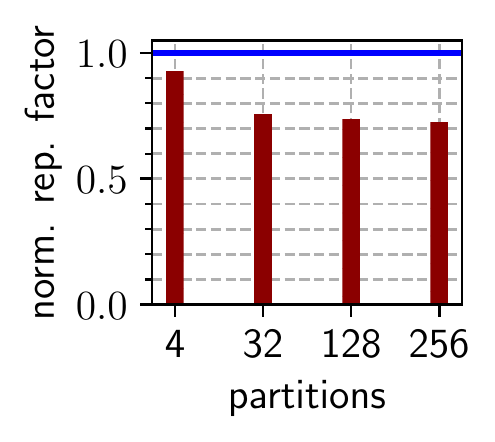}}
	\subfloat[FR: Run-time.]{\label{b}   \includegraphics[width=0.12\textwidth]{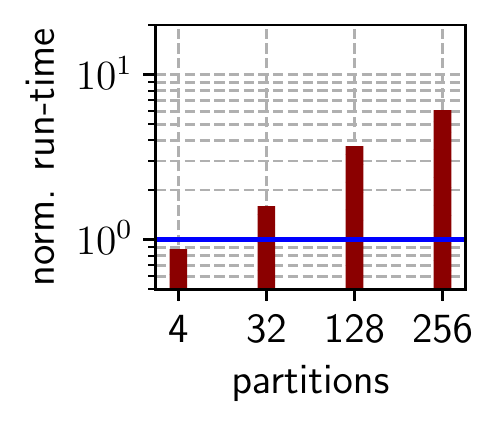}}
	\vspace{-2pt}
	\caption{Performance of 2PS-HDRF, normalized to the results of 2PS-L (blue line). }
	\label{eval:2ps-hdrf}
	\vspace{-12pt}
\end{figure*}

(2) 2PS-L yields a comparably good replication factor. In most cases, 2PS-L yields a lower replication factor than HDRF and ADWISE, which are stateful streaming partitioners that have much higher run-time (see discussion above). While in-memory partitioning (HEP, NE, DNE, METIS) still yields a better replication factor than 2PS-L, these partitioners also have a higher run-time and higher memory overhead. When comparing the partitioning quality of 2PS-L with DBH---the only partitioner that is continuously faster,---we see that 2PS-L yields significantly better replication factors on all graphs except for TW. The highest advantage of 2PS-L over DBH is on the GSH graph, where at $k=256$, the replication factor of DBH is $6.4 \times$ higher.

In summary, 2PS-L as an out-of-core edge partitioner shows superior performance to stateful streaming edge partitioning. We can reduce both replication factor \emph{and} run-time at the same time. Therefore, 2PS-L is an attractive new choice for out-of-core graph partitioning when both partitioning quality \emph{and} run-time are critical.

\subsection{Run-Time of Different Phases in 2PS-L.} 
\label{sec:runtime}
In Figure~\ref{eval:phases}, we dissect the total run-time of 2PS-L into its two phases, i.e., clustering and partitioning, and also report the time for calculating the vertex degrees. Between $7$ and $20~\%$ of the run-time are spent on degree calculations. This time could be saved if the vertex degrees are already known (which may be the case in practice, depending on the data source and format). Clustering takes between $16$ and $22~\%$ of the run-time. This time will increase when more streaming clustering passes are performed (see Section~\ref{sec:restreaming}). Finally, between $58$ and $77~\%$ of the run-time are accounted for in the partitioning phase, which includes the assignment of clusters to partitions, the pre-partitioning and the scoring-based partitioning pass.   

We see similar patterns in the distribution of run-time between the degree calculation, clustering and partitioning phases among the group of social network graphs (OK, TW, FR) and web graphs (IT, UK, GSH, WDC). This correlates to the ratio of the two different parts of the streaming partitioning phase, i.e., prepartitioning (assigning edges of commonly placed clusters to the single candidate partition) and partitioning of remaining edges (using the scoring function to decide between two candidate partitions). We show this ratio for the evaluated graphs in Figure~\ref{eval:edge_ratio}. Different from social network graphs, prepartitioning dominates in web graphs, which is faster than scoring-based partitioning although it has the same algorithmic time complexity. Therefore, web graphs exhibit a lower overall partitioning time and a lower fraction of the run-time is associated with partitioning.

\subsection{Re-Streaming}
\label{sec:restreaming}

We further evaluate how the partitioning quality is improved by re-streaming in the clustering phase of 2PS-L. We measured the relative gain in replication factor as compared to single-pass clustering in Figure~\ref{eval:restreaming_rf} as well as the run-time in Figure~\ref{eval:restreaming_runtime}. In terms of replication factor, the gains for re-streaming clustering are somewhat limited (up to 3.5~\% reduction). This needs to be paid for by a larger run-time; however, the increase in run-time is not proportional to the number of streaming passes. For example, for 8 streaming passes, the run-time roughly doubles as compared to single-pass clustering. This is because clustering only takes a minor portion of the total partitioning run-time (cf. Section~\ref{sec:runtime}). In the end, it depends on the concrete use cases to decide whether re-streaming pays off. Our recommended standard setting for 2PS-L is to perform a single streaming clustering pass, i.e., not apply re-streaming. Different from existing methods to reduce replication factors in out-of-core edge partitioning (e.g., ADWISE~\cite{8416335}), 2PS-L with re-streaming has a run-time that is independent of the number of partitions, so that the cost of re-streaming is still moderate compared to prior approaches.

\subsection{Comparison to HDRF Scoring}
\label{sec:hdrf_scoring}

We implemented an alternative version of 2PS-L that employs ``traditional'' stateful streaming partitioning with the HDRF scoring function~\cite{Petroni:2015:HSP:2806416.2806424} in the second phase instead of linear-time streaming partitioning as in 2PS-L. We call this version 2PS-HDRF. In the following, we compare the replication factor and the run-time of 2PS-HDRF with 2PS-L (see Figure~\ref{eval:2ps-hdrf}).

Using the HDRF scoring function improves the replication factor by up to $50~\%$. However, it comes at the cost of higher run-time with increasing number of partitions as a score is computed for every edge on every partition. At $k=4$, there is almost no run-time difference between 2PS-L and 2PS-HDRF. But at $k=256$, 2PS-L is up to $12 \times$ faster than 2PS-HDRF. Our recommendation is as follows: For a low number of partitions (like $k=4$), it pays off to use 2PS-HDRF, as the run-time is similar to 2PS-L, but the replication factor is lower. For a higher number of partitions ($k > 4$), the question of whether to use 2PS-L or 2PS-HDRF depends on the subsequent graph processing workload. It may pay off to invest more run-time into graph partitioning to yield a better partitioning quality that leads to faster graph processing, but it requires profiling of the graph processing performance to determine whether this is indeed the case. In Section~\ref{sec:processing}, we study the performance of graph processing under different partitionings to shed more light onto this question.

\subsection{Distributed Graph Processing}
\label{sec:processing}
We evaluate the distributed graph processing performance under different graph partitionings. To this end, we set up a cluster of 8 machines on which we equally distribute 32 Spark executors; details can be found in the appendix. We deploy Spark/GraphX version 3.0.0 and use static PageRank (PR) with 100 iterations as graph processing workload. We partitioned the graphs into the respective number of executors used, i.e., $k=32$. As baselines, we used state-of-the-art streaming edge partitioners, as well as HEP-1 (i.e., HEP with $\tau = 1.0$) which is an out-of-core partitioner that comes close to stateful streaming partitioners in terms of memory overhead~\cite{hep}.
Due to memory overheads in Spark (which internally uses property graphs represented as resilient distributed datasets), we could not process large graphs with more than one billion edges on our cluster. Therefore, we perform graph processing experiments on the OK graph (see Table~\ref{tab:graphs}) and a Wikipedia graph (WI) with 14 M vertices and 437 M edges~\cite{Kunegis:2013:KKN:2487788.2488173}.

\begin{table}
\scriptsize
	\begin{center}
		\begin{tabular}{|p{1.2cm}||l|l||l|l||l|l||l|l|}
			\hline
			\emph{Algor. /} &  \multicolumn{2}{c||}{Rep. Factor}  &  \multicolumn{2}{c||}{Partitioning} & \multicolumn{2}{c||}{PageRank}  & \multicolumn{2}{c|}{Total}   \\	
			\emph{Graph} & OK &  WI & OK &  WI  & OK  & WI & OK & WI \\	
			\hline \hline
			2PS-L & 9.00 & 4.55 & 20 & 80 & 240 &   786 & \textbf{260} &  \textbf{866} \\ \hline
			2PS-HDRF & 7.04 & 2.78 &  50 & 166 & \textbf{228}  &  730 & 278  & 896    \\ \hline \hline
			HDRF & 10.78 & 3.98 & 52 & 220 & 246  &  769 & 298 & 989  \\ \hline
			DBH & 12.42 & 5.72 & \textbf{6} & \textbf{28} & 285 	 & FAIL  & 291 & FAIL   \\ \hline
			SNE & 4.57 &  \textbf{2.21} & 110 & 574 & 230 &  \textbf{621} & 340 & 1,195   \\ \hline
			HEP-1 & \textbf{4.52}  & 2.59 & 45 & 244 & 261 &  632 & 306  & 876 \\ \hline
		\end{tabular} 
	\end{center}
\caption{Partitioning time and graph processing time (in seconds).}
\label{tab:processing}
\vspace{-14pt}
\end{table}

Table~\ref{tab:processing} shows the resulting replication factor, partitioning run-time and graph processing run-time (average of 3 runs). Neither the best partitioning quality (HEP-1 on OK, SNE on WI) nor the fastest partitioning (DBH) resulted in the best total run-time. HEP-1 and SNE yield the best replication factors, but as their partitioning run-time is relatively high, they do not perform best in an end-to-end comparison. On the other hand, DBH is the fastest partitioner. However, it also yields the worst replication factors, which makes graph processing slower. Therefore, in an end-to-end comparison, DBH does not perform best either. Even worse, due to a large replication factor on the WI graph, Spark/GraphX could not perform graph processing when the graph was partitioned with DBH. Instead, it ran out of disk space (35~GB per worker machine), as too much shuffling occurred due to the high replication factor.

We conclude that there \emph{is} a need for good partitioning quality, but partitioning run-time is equally important. 2PS-L takes into account both factors, as it is fast and yields a good replication factor at the same time. As a consequence, the total run-time was always lowest when partitioning the graph with 2PS-L. 2PS-HDRF achieved a better replication factor and a lower graph processing run-time; however, due to its higher partitioning run-time, it did not perform better in terms of total run-time.

\begin{table}
\scriptsize
\setlength{\tabcolsep}{0.5em}
	\begin{center}
		\begin{tabular}{|p{1.2cm}|p{0.8cm}|p{0.8cm}|p{0.8cm}|p{0.8cm}|p{0.8cm}|p{0.8cm}|p{0.8cm}|}
			\hline
			 & OK &  IT & TW &  FR  & UK  & GSH & WDC  \\	
			\hline \hline
			Page Cache & 24 s 						& 96 s 						& 	7.3 m					 & 15.4 m			 			& 5.3 m 				 			&  69 m 						 		& 131 m  \\ \hline
			SSD & 29 s \newline +22 \% & 135 s \newline +40 \% & 8.2 m \newline +12 \% & 16.5 m \newline +7 \% & 7.1 m \newline +34 \%  &  78 m 	\newline +13 \%	 & 149 m \newline +14 \%     \\ \hline
			HDD & 61 s \newline +159 \% & 393 s \newline +308 \% & 14.2 m \newline +93 \% & 23.8 m \newline +54 \% & 20.5 m \newline +285 \%  &  206 m \newline +200 \% & 411 m \newline +214 \%  \\ \hline
		\end{tabular} 
	\end{center}
\caption{Partitioning time using different storage devices.}
\label{tab:external}
\vspace{-12pt}
\end{table}

\subsection{External Storage}
\label{sec:external}

Loading the graph data from external storage may slow down the performance of 2PS-L, in particular, as multiple passes through the edge list need to be performed. To evaluate the impact of I/O on partitioning time, we set up a server with two different storage devices: a local SSD and a local HDD. We profiled the sequential read speed using \texttt{fio} (single-threaded reading of a single file of 5 GB size in 100 MB blocks, average of 3 runs), resulting in 938 MB/s for the SSD and 158 MB/s for the HDD. To force 2PS-L to read the graph in every streaming iteration completely from disk, we drop the page cache~\cite{pagecache} after each streaming pass which invalidates the cached disk blocks for subsequent streaming passes. In Table~\ref{tab:external}, we compare the partitioning run-time under different storage solutions (page cache, SSD, HDD). The SSD is between 7~\% and 22~\% slower on social network graphs, and between 13~\% and 40~\% slower on web graphs, respectively, compared to reading the graph data from the page cache. Using an HDD comes with a performance penalty of 54~\% to 159~\% on social network graphs and 200~\% to 308~\% on web graphs, respectively. In conclusion, we recommend to employ a fast storage that achieves at least 1 GB/s sequential read speed when using 2PS-L in memory-constrained situations where none of the graph data can be cached in memory.

%-------------------------------------------------------------------------------
%-------------------------------------------------------------------------------
\section{Related Work}
\label{sec:related}

Graph partitioning is a problem with a long history in research~\cite{gp-survey}. It has numerous applications in solving optimization problems, e.g., in VLSI design~\cite{Karypis:1998:FHQ:305219.305248}, operator placement in stream processing systems~\cite{10.1007/978-3-642-10445-9_16} and data management in graph-based data analytics systems ~\cite{powergraph, graphx, 10.14778/2212351.2212354, 10.1145/2384716.2384777}.  For both vertex partitioning and edge partitioning, in-memory partitioners as well as out-of-core partitioners have been proposed. 

In-memory partitioners load the complete graph into memory of either a single machine or a cluster of multiple machines before partitioning. A classical approach to in-memory partitioning are \emph{multi-level partitioners} such as SCOTCH~\cite{10.5555/645560.658570},  METIS~\cite{Karypis:1998:FHQ:305219.305248} or KaHiP~\cite{schlag2019scalable}. They work on the principle of coarsening the graph, splitting the coarsened graph, and then refining the splitting to the original graph which results in the final partitioning. While 2PS-L shares the idea of building an intermediary graph representation (i.e., the vertex clustering), it does not perform partitioning on different levels, but instead only partitions the graph once in the final streaming phase. Other in-memory partitioners build different intermediary data structures, e.g., elimination trees~\cite{Margo:2015:SDG:2824032.2824046} or split graphs~\cite{schlag2019scalable}, or they materialize the original graph in memory using variants of the compressed sparse row representation~\cite{Zhang:2017:GEP:3097983.3098033, dne}. None of these intermediary data structures can easily be built in a streaming fashion, and we wanted to avoid an in-memory materialization of the complete graph. TLP~\cite{TLP} is a semi-streaming algorithm that consists of multiple phases; however, the unpartitioned graph data needs to be repeatedly sorted and streamed in different breadth-first-search orderings, which induces prohibitively large overhead. 

Stanton and Kliot~\cite{Stanton:2012:SGP:2339530.2339722} proposed a streaming model for vertex partitioning that is also used in the FENNEL~\cite{Tsourakakis:2014:FSG:2556195.2556213} partitioner. In this model, different from streaming edge partitioning, \emph{adjacency lists} are ingested and vertices are assigned to partitions with the goal of minimizing the number of cut edges. Awadelkarim and Ugander~\cite{uganderkdd20} analyze the impact of the stream order on streaming vertex partitioning; this could also be interesting for streaming clustering. In streaming edge partitioning, besides stateless partitioning based on hashing (e.g., DBH~\cite{dbh} and Grid~\cite{grid}), stateful partitioners have received growing attention. Here, by gathering state about past assignments, future edges in the stream can be partitioned with a lower replication factor. Different scoring functions have been proposed, e.g., Greedy~\cite{powergraph} or HDRF~\cite{Petroni:2015:HSP:2806416.2806424}, but their time complexity is linear in the number of edges \emph{and} number of partitions (i.e., $\mathcal{O}(|E| * k)$), making them expensive. Further, these approaches suffer from the uninformed assignment problem. To overcome this problem, ADWISE~\cite{8416335} introduced a buffer for re-ordering edges. However, ADWISE only considers edges \emph{within the buffer}, which does not work well for very large graphs. Different from that, 2PS-L pre-processes the entire graph and, hence, has a \emph{global} view of the graph structure. Hence, 2PS-L is effective on graphs of any size. RBSEP~\cite{taimouri2019rbsep} in this sense is similar to ADWISE, as it also relies on buffering edges and thus, suffers from the same limitations. Quasi-streaming~\cite{8598984} is a heuristic that consumes the edge stream as a set of non-overlapping chunks (e.g., 6,400 edges per chunk) and assigns these edges to partitions based on a game-theoretical problem formulation. This way, memory overhead can be kept low while the achieved replication factor is better than with pure streaming algorithms. However, quasi-streaming has a large run-time, which is comparable to METIS, while 2PS-L is up to several orders of magnitude faster than METIS. Another approach to out-of-core edge partitioning is Hybrid Edge Partitioner (HEP)~\cite{hep}, a hybrid between in-memory and streaming partitioning. However, HEP uses the HDRF scoring function in its streaming phase, thus leading to high run-time of streaming partitioning.

Some of the edge partitioning algorithms in the literature report theoretical bounds on the replication factor~\cite{Petroni:2015:HSP:2806416.2806424, Zhang:2017:GEP:3097983.3098033, dbh}. These bounds commonly rely on the assumption that the input graph strictly follows a power-law degree distribution. However, not all real-world graphs are ``true'' power-law graphs, which has unforeseeable consequences on the validity of the bounds. A further problem is that the community structure of graphs is neglected in those bounds, although it plays a major role in graph partitioning. In particular, analysis techniques from other graph partitioning approaches to determine quality bounds are not directly applicable to 2PS-L~\cite{Petroni:2015:HSP:2806416.2806424, Zhang:2017:GEP:3097983.3098033, dbh}, as the result of the clustering phase influences the quality of the partitioning phase. In conclusion, we think that there needs to be more work on the theoretical foundations of partitioning quality before such bounds become practical for assessing the real-world performance of partitioners. Instead of relying on theoretical bounds, we compared the replication factor of 2PS-L to its competitors experimentally on a large set of real-world graphs which provides a realistic picture of what to expect at real workloads.

While we focus on devising an efficient out-of-core edge partitioning algorithm, other works extend the concept of streaming partitioning to support dynamic graphs and parallelization. Following the approach proposed by Fan et al.~\cite{incremental}, 2PS-L could be transformed into an incremental algorithm to efficiently handle dynamic graphs with edge insertions and deletions without recomputing the complete partitioning from scratch. Li et al.~\cite{9388864} propose a complementary approach to re-assign edges to partitions under dynamic graph updates. CuSP~\cite{cusp} is a framework for parallelizing streaming edge partitioning algorithms. While our focus is not on parallelization but on reducing time complexity, 2PS-L could be integrated into the CuSP framework to speed up the partitioning. However, parallelization comes with a cost, as staleness in state synchronization of multiple partitioner instances can lead to lower partitioning quality. Fan et al.~\cite{10.1145/3318464.3389745} propose a machine learning based approach to refine a given partitioning to better suit a given graph processing problem. In this respect, 2PS-L can be used as an efficient initial partitioner.

Graph partitioning is tightly related to the \emph{vertex separator} problem, i.e., finding a set of vertices that split a graph into equally-sized components. Brandt and Wattenhofer~\cite{vertex_separator} propose a random Las Vegas algorithm. However, their work has not been implemented and evaluated on real workloads, does not exploit power-law degree distributions or clustering and does not suit the out-of-core setting that we target.

There are many graph clustering algorithms, such as label propagation~\cite{raghavan2007near} or incremental aggregation~\cite{shiokawa2013fast}. We base our work on Hollocou's algorithm~\cite{hollocou2017streaming}, as it only induces little memory overheads and has linear run-time, but other clustering algorithms could possibly be integrated with \mbox{2PS-L}.

Some graph processing systems employ special execution models that require a specific style of graph pre-processing. PowerLyra~\cite{powerlyra} is a hybrid distributed graph processing system that uses a combination of vertex-centric and edge-centric processing. Therefore, it also employs its own hybrid partitioning strategy that combines vertex partitioning and edge partitioning.  Gemini~\cite{199295} performs chunk-based partitioning of vertices, exploiting locality within the graph (neighboring vertices being expected to have similar IDs). 

Out-of-core graph processing systems like GraphChi~\cite{graphchi}, GridGraph~\cite{gridgraph} and Mosaic~\cite{Maass:2017:MPT:3064176.3064191} employ different graph pre-processing methods that involve specific graph partitioning problems. While these systems perform graph computations using external memory, they are not tailored to efficiently solve the edge partitioning problem.

A related problem is hypergraphs partitioning. In hypergraphs, an edge (called hyperedge) can connect more than two vertices. This way, group relationships can be modeled. Hypergraph partitioning has also been tackled by streaming~\cite{Alistarh:2015:SMH:2969442.2969452} as well as in-memory algorithms~\cite{1639359, TRIFUNOVIC2008563, Kabiljo:2017:SHP:3137628.3137650, 8621968}.

%-------------------------------------------------------------------------------%-------------------------------------------------------------------------------
\section{Conclusions}
\label{sec:conclusions}
In this paper, we propose the new two-phase out-of-core algorithm 2PS-L for edge partitioning of large graphs. We make use of the great flexibility of graph clustering in the first phase before we finalize and materialize the actual partitioning in the second phase. This way, we achieve state-of-the-art results in replication factor while maintaining low space complexity. Furthermore, we introduce a new linear-time scoring mechanism and function in 2PS-L, which reduces time complexity.
Finally, we show that partitioning with 2PS-L can lead to improved end-to-end runtime of partititoning plus distributed graph processing. 
In future work, we plan to investigate the generalization of 2PS-L to hypergraphs.

%-------------------------------------------------------------------------------

\section*{Appendix: Experimental Settings}
\label{sec:appendix}

Experiments in Section V.A--E are performed on a server with 4 x 8 Intel(R) Xeon(R) CPU E5-46500@2.70GHz and 528 GB of memory with Ubuntu 18.04.2 LTS; in Section V.F, we used a server with 4 x 14 Intel(R) Xeon(R) CPU E7-4850 v3 @ 2.20GHz and 528 GB of memory with Ubuntu 20.04.2 LTS. To evaluate the distributed graph processing performance in Section V.E, we set up a cluster of 4 compute nodes with 10 GBit Ethernet links, each having 41 GiB of RAM and 16 CPU cores, hosted on an OpenStack private compute cloud.

We configured the system parameters of the baseline partitioners according to the respective authors' recommendations as follows. For HDRF, we set $\lambda = 1.1$. For SNE, we used a cache size of $2*|V|$. For DNE, we set an expansion ratio of 0.1. For 2PS-L, we perform only one streaming pass in the clustering phase (no re-streaming).

We implemented 2PS-L as a separate process that reads the graph data as a file from a given storage, partitions the edges, and writes back the partitioned graph data to storage. This partitioned graph data can then be ingested by a data loader into the data processing framework of choice for subsequent processing. For a fair comparison, we re-implemented the HDRF algorithm in C++ and use the same implementation of HDRF also for edge partitioning in the second phase of 2PS-HDRF. We also re-implemented DBH, using the same framework that we developed for HDRF and 2PS-L. For NE, SNE, DNE, HEP and ADWISE, we use the reference implementation of the respective authors. All implementations that we use, except for DNE, METIS and ADWISE, ingest the graph in the same binary input format (i.e., binary edge list with 32-bit vertex IDs). DNE, METIS and ADWISE ingest ASCII edge or adjacency lists due to their implementation. However, the graph ingestion time of these systems is still negligible, as they either have a large total run-time (ADWISE and METIS) or perform parallel ingestion (DNE).

For each experiment, we re-compiled the partitioners with optimal settings. For 2PS, HDRF and DNE, we set the maximum number of partitions to the $k$ used in the corresponding experiment; this optimizes their respective memory overheads. Further, we compiled DNE with 32-bit vertex IDs, which minimizes its memory overhead; all other partitioners use 32-bit vertex IDs already by default.
In DNE, for each partition, a separate process is spawned. As our evaluation machine offers 64 hardware threads, we assigned each process $\lceil\frac{64}{k}\rceil$ threads.

Different from the other partitioners, for ADWISE and METIS we only performed each partitioning experiment once, as they take a lot of time. For each set of experiments, we perform an initial warm-up run which does not count toward the results. To use our limited server time efficiently, we aborted experiments after 12 hours for the smaller graphs with less than 10~B edges, except for some runs of ADWISE and METIS which we let run longer in order to gather enough data points. Due to long run-times, experiments on GSH and WDC graphs were performed only once.

ADWISE allows for setting a run-time bound that controls the buffer size, and thus, the partitioning quality. To get the best possible result in terms of replication factor, we set the run-time bound of ADWISE generously to 40 times the run-time of HDRF (which it did not keep at all times). Furthermore, we set a minimum buffer size of 10 edges in order to force ADWISE to still use a buffer even if the run-time bound would be exceeded. Under these favorable conditions, ADWISE was able to reach better replication factors than HDRF on the smaller graphs (OK and IT). However, on the larger graphs, the buffer size of ADWISE still was not large enough to yield an improvement over HDRF. 

\textbf{Code availability.} Our source code is available online (\url{https://github.com/mayerrn/two_phase_streaming}). We also provide links to all datasets that we have used in our paper.

%\clearpage

%\bibliographystyle{ACM-Reference-Format}
\bibliographystyle{IEEEtran}
\IEEEtriggeratref{55}
\bibliography{paper}

% Generated by IEEEtran.bst, version: 1.14 (2015/08/26)
\begin{thebibliography}{10}
\providecommand{\url}[1]{#1}
\csname url@samestyle\endcsname
\providecommand{\newblock}{\relax}
\providecommand{\bibinfo}[2]{#2}
\providecommand{\BIBentrySTDinterwordspacing}{\spaceskip=0pt\relax}
\providecommand{\BIBentryALTinterwordstretchfactor}{4}
\providecommand{\BIBentryALTinterwordspacing}{\spaceskip=\fontdimen2\font plus
\BIBentryALTinterwordstretchfactor\fontdimen3\font minus
  \fontdimen4\font\relax}
\providecommand{\BIBforeignlanguage}[2]{{%
\expandafter\ifx\csname l@#1\endcsname\relax
\typeout{** WARNING: IEEEtran.bst: No hyphenation pattern has been}%
\typeout{** loaded for the language `#1'. Using the pattern for}%
\typeout{** the default language instead.}%
\else
\language=\csname l@#1\endcsname
\fi
#2}}
\providecommand{\BIBdecl}{\relax}
\BIBdecl

\bibitem{graphx}
J.~E. Gonzalez, R.~S. Xin, A.~Dave, D.~Crankshaw, M.~J. Franklin, and
  I.~Stoica, ``Graphx: Graph processing in a distributed dataflow framework,''
  in \emph{11th {USENIX} Symposium on Operating Systems Design and
  Implementation ({OSDI} 14)}.\hskip 1em plus 0.5em minus 0.4em\relax
  Broomfield, CO: {USENIX} Association, 2014, pp. 599--613.

\bibitem{powergraph}
J.~E. Gonzalez, Y.~Low, H.~Gu, D.~Bickson, and C.~Guestrin, ``Powergraph:
  Distributed graph-parallel computation on natural graphs,'' in \emph{10th
  {USENIX} Symposium on Operating Systems Design and Implementation ({OSDI}
  12)}.\hskip 1em plus 0.5em minus 0.4em\relax Hollywood, CA: {USENIX}, 2012,
  pp. 17--30.

\bibitem{pregel}
G.~Malewicz, M.~H. Austern, A.~J. Bik, J.~C. Dehnert, I.~Horn, N.~Leiser, and
  G.~Czajkowski, ``Pregel: A system for large-scale graph processing,'' in
  \emph{Proceedings of the 2010 ACM SIGMOD International Conference on
  Management of Data}, ser. SIGMOD '10.\hskip 1em plus 0.5em minus 0.4em\relax
  New York, NY, USA: ACM, 2010, pp. 135--146.

\bibitem{giraph}
\BIBentryALTinterwordspacing
{N.N.}, ``{Apache Giraph}.'' [Online]. Available:
  \url{https://giraph.apache.org/}
\BIBentrySTDinterwordspacing

\bibitem{10.14778/2212351.2212354}
Y.~Low, D.~Bickson, J.~Gonzalez, C.~Guestrin, A.~Kyrola, and J.~M. Hellerstein,
  ``Distributed graphlab: A framework for machine learning and data mining in
  the cloud,'' \emph{Proc. VLDB Endow.}, vol.~5, no.~8, p. 716–727, Apr.
  2012.

\bibitem{10.1145/2384716.2384777}
J.~Webber, ``A programmatic introduction to neo4j,'' in \emph{Proceedings of
  the 3rd Annual Conference on Systems, Programming, and Applications: Software
  for Humanity}, ser. SPLASH ’12.\hskip 1em plus 0.5em minus 0.4em\relax New
  York, NY, USA: ACM, 2012, p. 217–218.

\bibitem{dgl}
{N.N.}, ``Deep graph library,'' https://www.dgl.ai/, 2021.

\bibitem{MLSYS2020_fe9fc289}
Z.~Jia, S.~Lin, M.~Gao, M.~Zaharia, and A.~Aiken, ``Improving the accuracy,
  scalability, and performance of graph neural networks with roc,'' in
  \emph{Proceedings of Machine Learning and Systems}, I.~Dhillon,
  D.~Papailiopoulos, and V.~Sze, Eds., vol.~2, 2020, pp. 187--198.

\bibitem{Bourse:2014:BGE:2623330.2623660}
F.~Bourse, M.~Lelarge, and M.~Vojnovic, ``Balanced graph edge partition,'' in
  \emph{Proceedings of the 20th ACM SIGKDD International Conference on
  Knowledge Discovery and Data Mining}, ser. KDD '14.\hskip 1em plus 0.5em
  minus 0.4em\relax New York, NY, USA: ACM, 2014, pp. 1456--1465.

\bibitem{Zhang:2017:GEP:3097983.3098033}
C.~Zhang, F.~Wei, Q.~Liu, Z.~G. Tang, and Z.~Li, ``Graph edge partitioning via
  neighborhood heuristic,'' in \emph{Proceedings of the 23rd ACM SIGKDD
  International Conference on Knowledge Discovery and Data Mining}, ser. KDD
  '17.\hskip 1em plus 0.5em minus 0.4em\relax New York, NY, USA: ACM, 2017, pp.
  605--614.

\bibitem{Karypis:1998:FHQ:305219.305248}
G.~Karypis and V.~Kumar, ``A fast and high quality multilevel scheme for
  partitioning irregular graphs,'' \emph{SIAM J. Sci. Comput.}, vol.~20, no.~1,
  pp. 359--392, Dec. 1998.

\bibitem{schlag2019scalable}
S.~Schlag, C.~Schulz, D.~Seemaier, and D.~Strash, ``Scalable edge
  partitioning,'' in \emph{2019 Proceedings of the Twenty-First Workshop on
  Algorithm Engineering and Experiments (ALENEX)}.\hskip 1em plus 0.5em minus
  0.4em\relax SIAM, 2019, pp. 211--225.

\bibitem{Margo:2015:SDG:2824032.2824046}
D.~Margo and M.~Seltzer, ``A scalable distributed graph partitioner,''
  \emph{Proc. VLDB Endow.}, vol.~8, no.~12, pp. 1478--1489, Aug. 2015.

\bibitem{dne}
M.~Hanai, T.~Suzumura, W.~J. Tan, E.~Liu, G.~Theodoropoulos, and W.~Cai,
  ``Distributed edge partitioning for trillion-edge graphs,'' \emph{Proc. VLDB
  Endow.}, vol.~12, no.~13, p. 2379–2392, Sep. 2019.

\bibitem{Stanton:2012:SGP:2339530.2339722}
I.~Stanton and G.~Kliot, ``Streaming graph partitioning for large distributed
  graphs,'' in \emph{Proceedings of the 18th ACM SIGKDD International
  Conference on Knowledge Discovery and Data Mining}, ser. KDD '12.\hskip 1em
  plus 0.5em minus 0.4em\relax New York, NY, USA: ACM, 2012, pp. 1222--1230.

\bibitem{dbh}
C.~Xie, L.~Yan, W.-J. Li, and Z.~Zhang, ``Distributed power-law graph
  computing: Theoretical and empirical analysis,'' in \emph{Advances in Neural
  Information Processing Systems 27}, 2014, pp. 1673--1681.

\bibitem{Petroni:2015:HSP:2806416.2806424}
F.~Petroni, L.~Querzoni, K.~Daudjee, S.~Kamali, and G.~Iacoboni, ``Hdrf:
  Stream-based partitioning for power-law graphs,'' in \emph{Proceedings of the
  24th ACM International on Conference on Information and Knowledge
  Management}, ser. CIKM '15.\hskip 1em plus 0.5em minus 0.4em\relax New York,
  NY, USA: ACM, 2015, pp. 243--252.

\bibitem{8416335}
C.~Mayer, R.~Mayer, M.~A. Tariq, H.~Geppert, L.~Laich, L.~Rieger, and
  K.~Rothermel, ``Adwise: Adaptive window-based streaming edge partitioning for
  high-speed graph processing,'' in \emph{2018 IEEE 38th International
  Conference on Distributed Computing Systems (ICDCS)}, July 2018, pp.
  685--695.

\bibitem{hep}
R.~Mayer and H.-A. Jacobsen, ``Hybrid edge partitioner: Partitioning large
  power-law graphs under memory constraints,'' in \emph{Proceedings of the 2021
  International Conference on Management of Data}, ser. SIGMOD/PODS '21.\hskip
  1em plus 0.5em minus 0.4em\relax New York, NY, USA: Association for Computing
  Machinery, 2021, p. 1289–1302.

\bibitem{Pacaci:2019:EAS:3299869.3300076}
A.~Pacaci and M.~T. \"{O}zsu, ``Experimental analysis of streaming algorithms
  for graph partitioning,'' in \emph{Proceedings of the 2019 International
  Conference on Management of Data}, ser. SIGMOD '19.\hskip 1em plus 0.5em
  minus 0.4em\relax New York, NY, USA: ACM, 2019, pp. 1375--1392.

\bibitem{Abbas:2018:SGP:3236187.3269471}
Z.~Abbas, V.~Kalavri, P.~Carbone, and V.~Vlassov, ``Streaming graph
  partitioning: An experimental study,'' \emph{Proc. VLDB Endow.}, vol.~11,
  no.~11, pp. 1590--1603, Jul. 2018.

\bibitem{verma-vldb}
S.~Verma, L.~M. Leslie, Y.~Shin, and I.~Gupta, ``An experimental comparison of
  partitioning strategies in distributed graph processing,'' \emph{Proc. VLDB
  Endow.}, vol.~10, no.~5, pp. 493--504, Jan. 2017.

\bibitem{10.1145/3434642}
S.~Sakr, A.~Bonifati, H.~Voigt, A.~Iosup, K.~Ammar, R.~Angles, W.~Aref,
  M.~Arenas, M.~Besta, P.~A. Boncz, K.~Daudjee, E.~D. Valle, S.~Dumbrava,
  O.~Hartig, B.~Haslhofer, T.~Hegeman, J.~Hidders, K.~Hose, A.~Iamnitchi,
  V.~Kalavri, H.~Kapp, W.~Martens, M.~T. \"{O}zsu, E.~Peukert, S.~Plantikow,
  M.~Ragab, M.~R. Ripeanu, S.~Salihoglu, C.~Schulz, P.~Selmer, J.~F. Sequeda,
  J.~Shinavier, G.~Sz\'{a}rnyas, R.~Tommasini, A.~Tumeo, A.~Uta, A.~L.
  Varbanescu, H.-Y. Wu, N.~Yakovets, D.~Yan, and E.~Yoneki, ``The future is big
  graphs: A community view on graph processing systems,'' \emph{Commun. ACM},
  vol.~64, no.~9, p. 62–71, Aug. 2021.

\bibitem{gandhi2021p3}
S.~Gandhi and A.~P. Iyer, ``P3: Distributed deep graph learning at scale,'' in
  \emph{15th $\{$USENIX$\}$ Symposium on Operating Systems Design and
  Implementation ($\{$OSDI$\}$ 21)}, 2021, pp. 551--568.

\bibitem{Newman8577}
M.~E.~J. Newman, ``Modularity and community structure in networks,''
  \emph{Proceedings of the National Academy of Sciences}, vol. 103, no.~23, pp.
  8577--8582, 2006.

\bibitem{FORTUNATO201075}
S.~Fortunato, ``Community detection in graphs,'' \emph{Physics Reports}, vol.
  486, no.~3, pp. 75 -- 174, 2010.

\bibitem{hollocou2017streaming}
A.~Hollocou, J.~Maudet, T.~Bonald, and M.~Lelarge, ``A streaming algorithm for
  graph clustering,'' in \emph{NIPS 2017-Wokshop on Advances in Modeling and
  Learning Interactions from Complex Data}, 2017, pp. 1--10.

\bibitem{Nishimura:2013:RGP:2487575.2487696}
J.~Nishimura and J.~Ugander, ``Restreaming graph partitioning: Simple versatile
  algorithms for advanced balancing,'' in \emph{Proceedings of the 19th ACM
  SIGKDD International Conference on Knowledge Discovery and Data Mining}, ser.
  KDD '13.\hskip 1em plus 0.5em minus 0.4em\relax New York, NY, USA: ACM, 2013,
  pp. 1106--1114.

\bibitem{graham1969bounds}
R.~L. Graham, ``Bounds on multiprocessing timing anomalies,'' \emph{SIAM
  journal on Applied Mathematics}, vol.~17, no.~2, pp. 416--429, 1969.

\bibitem{Ullman:1975:NSP:1739944.1740138}
J.~D. Ullman, ``Np-complete scheduling problems,'' \emph{J. Comput. Syst.
  Sci.}, vol.~10, no.~3, pp. 384--393, Jun. 1975.

\bibitem{6413740}
J.~Yang and J.~Leskovec, ``Defining and evaluating network communities based on
  ground-truth,'' in \emph{2012 IEEE 12th International Conference on Data
  Mining}, Dec 2012, pp. 745--754.

\bibitem{snapnets}
J.~Leskovec and A.~Krevl, ``{SNAP Datasets}: {Stanford} large network dataset
  collection,'' http://snap.stanford.edu/data, Jun. 2014.

\bibitem{BMSB}
P.~Boldi, A.~Marino, M.~Santini, and S.~Vigna, ``{BUbiNG}: Massive crawling for
  the masses,'' in \emph{Proceedings of the Companion Publication of the 23rd
  International Conference on World Wide Web}.\hskip 1em plus 0.5em minus
  0.4em\relax ACM, 2014, pp. 227--228.

\bibitem{BRSLLP}
P.~Boldi, M.~Rosa, M.~Santini, and S.~Vigna, ``Layered label propagation: A
  multiresolution coordinate-free ordering for compressing social networks,''
  in \emph{Proceedings of the 20th international conference on World Wide
  Web}.\hskip 1em plus 0.5em minus 0.4em\relax ACM Press, 2011, pp. 587--596.

\bibitem{BoVWFI}
P.~Boldi and S.~Vigna, ``The {W}eb{G}raph framework {I}: {C}ompression
  techniques,'' in \emph{Proc. of the Thirteenth International World Wide Web
  Conference (WWW 2004)}.\hskip 1em plus 0.5em minus 0.4em\relax Manhattan,
  USA: ACM, 2004, pp. 595--601.

\bibitem{wdc}
\BIBentryALTinterwordspacing
{N.N.}, ``{WDC} graph.'' [Online]. Available:
  \url{http://webdatacommons.org/hyperlinkgraph/}
\BIBentrySTDinterwordspacing

\bibitem{grid}
N.~Jain, G.~Liao, and T.~L. Willke, ``Graphbuilder: Scalable graph etl
  framework,'' in \emph{First International Workshop on Graph Data Management
  Experiences and Systems}, ser. GRADES '13.\hskip 1em plus 0.5em minus
  0.4em\relax New York, NY, USA: ACM, 2013, pp. 4:1--4:6.

\bibitem{spinner}
C.~Martella, D.~Logothetis, A.~Loukas, and G.~Siganos, ``Spinner: Scalable
  graph partitioning in the cloud,'' in \emph{2017 IEEE 33rd International
  Conference on Data Engineering (ICDE)}, April 2017, pp. 1083--1094.

\bibitem{10.1145/369028.369103}
G.~Karypis and V.~Kumar, ``Parallel multilevel k-way partitioning scheme for
  irregular graphs,'' in \emph{Proceedings of the 1996 ACM/IEEE Conference on
  Supercomputing}, ser. Supercomputing ’96.\hskip 1em plus 0.5em minus
  0.4em\relax USA: IEEE Computer Society, 1996, p. 35–es.

\bibitem{xtrapulp}
G.~M. Slota, S.~Rajamanickam, K.~Devine, and K.~Madduri, ``Partitioning
  trillion-edge graphs in minutes,'' in \emph{2017 IEEE International Parallel
  and Distributed Processing Symposium (IPDPS)}, May 2017, pp. 646--655.

\bibitem{Kunegis:2013:KKN:2487788.2488173}
J.~Kunegis, ``Konect: The koblenz network collection,'' in \emph{Proceedings of
  the 22Nd International Conference on World Wide Web}, ser. WWW '13
  Companion.\hskip 1em plus 0.5em minus 0.4em\relax New York, NY, USA: ACM,
  2013, pp. 1343--1350.

\bibitem{pagecache}
\BIBentryALTinterwordspacing
``{The Linux Kernel documentation},'' last Accessed 02/2022. [Online].
  Available:
  \url{https://www.kernel.org/doc/html/latest/admin-guide/mm/concepts.html\#page-cache}
\BIBentrySTDinterwordspacing

\bibitem{gp-survey}
A.~Bulu{\c{c}}, H.~Meyerhenke, I.~Safro, P.~Sanders, and C.~Schulz,
  \emph{Recent Advances in Graph Partitioning}.\hskip 1em plus 0.5em minus
  0.4em\relax Cham: Springer International Publishing, 2016, pp. 117--158.

\bibitem{10.1007/978-3-642-10445-9_16}
R.~Khandekar, K.~Hildrum, S.~Parekh, D.~Rajan, J.~Wolf, K.-L. Wu, H.~Andrade,
  and B.~Gedik, ``Cola: Optimizing stream processing applications via graph
  partitioning,'' in \emph{Middleware 2009}, J.~M. Bacon and B.~F. Cooper,
  Eds.\hskip 1em plus 0.5em minus 0.4em\relax Berlin, Heidelberg: Springer
  Berlin Heidelberg, 2009, pp. 308--327.

\bibitem{10.5555/645560.658570}
F.~Pellegrini and J.~Roman, ``Scotch: A software package for static mapping by
  dual recursive bipartitioning of process and architecture graphs,'' in
  \emph{Proceedings of the International Conference and Exhibition on
  High-Performance Computing and Networking}, ser. HPCN Europe 1996.\hskip 1em
  plus 0.5em minus 0.4em\relax Berlin, Heidelberg: Springer-Verlag, 1996, p.
  493–498.

\bibitem{TLP}
S.~Ji, C.~Bu, L.~Li, and X.~Wu, ``Local graph edge partitioning with a
  two-stage heuristic method,'' in \emph{2019 IEEE 39th International
  Conference on Distributed Computing Systems (ICDCS)}, 2019, pp. 228--237.

\bibitem{Tsourakakis:2014:FSG:2556195.2556213}
C.~Tsourakakis, C.~Gkantsidis, B.~Radunovic, and M.~Vojnovic, ``Fennel:
  Streaming graph partitioning for massive scale graphs,'' in \emph{Proceedings
  of the 7th ACM International Conference on Web Search and Data Mining}, ser.
  WSDM '14.\hskip 1em plus 0.5em minus 0.4em\relax New York, NY, USA: ACM,
  2014, pp. 333--342.

\bibitem{uganderkdd20}
A.~Awadelkarim and J.~Ugander, ``Prioritized restreaming algorithms for
  balanced graph partitioning,'' in \emph{Proceedings of the 26th ACM SIGKDD
  International Conference on Knowledge Discovery and Data Mining}.\hskip 1em
  plus 0.5em minus 0.4em\relax New York, NY, USA: ACM, 2020, p. 1877–1887.

\bibitem{taimouri2019rbsep}
M.~Taimouri and H.~Saadatfar, ``Rbsep: a reassignment and buffer based
  streaming edge partitioning approach,'' \emph{Journal of Big Data}, vol.~6,
  no.~1, pp. 1--17, 2019.

\bibitem{8598984}
Q.-S. Hua, Y.~Li, D.~Yu, and H.~Jin, ``Quasi-streaming graph partitioning: A
  game theoretical approach,'' \emph{IEEE Transactions on Parallel and
  Distributed Systems}, vol.~30, no.~7, pp. 1643--1656, 2019.

\bibitem{incremental}
W.~Fan, M.~Liu, C.~Tian, R.~Xu, and J.~Zhou, ``Incrementalization of graph
  partitioning algorithms,'' \emph{Proc. VLDB Endow.}, vol.~13, no.~8, 2020.

\bibitem{9388864}
H.~Li, H.~Yuan, J.~Huang, J.~Cui, X.~Ma, S.~Wang, J.~Yoo, and P.~S. Yu, ``Group
  reassignment for dynamic edge partitioning,'' \emph{IEEE Transactions on
  Parallel and Distributed Systems}, vol.~32, no.~10, pp. 2477--2490, 2021.

\bibitem{cusp}
L.~Hoang, R.~Dathathri, G.~Gill, and K.~Pingali, ``Cusp: A customizable
  streaming edge partitioner for distributed graph analytics,'' in \emph{2019
  IEEE International Parallel and Distributed Processing Symposium (IPDPS)},
  May 2019, pp. 439--450.

\bibitem{10.1145/3318464.3389745}
\BIBentryALTinterwordspacing
W.~Fan, R.~Jin, M.~Liu, P.~Lu, X.~Luo, R.~Xu, Q.~Yin, W.~Yu, and J.~Zhou,
  ``Application driven graph partitioning,'' in \emph{Proceedings of the 2020
  ACM SIGMOD International Conference on Management of Data}, ser. SIGMOD
  '20.\hskip 1em plus 0.5em minus 0.4em\relax New York, NY, USA: Association
  for Computing Machinery, 2020, p. 1765–1779. [Online]. Available:
  \url{https://doi.org/10.1145/3318464.3389745}
\BIBentrySTDinterwordspacing

\bibitem{vertex_separator}
S.~Brandt and R.~Wattenhofer, ``Approximating small balanced vertex separators
  in almost linear time,'' in \emph{Algorithms and Data Structures}, F.~Ellen,
  A.~Kolokolova, and J.-R. Sack, Eds.\hskip 1em plus 0.5em minus 0.4em\relax
  Cham: Springer International Publishing, 2017, pp. 229--240.

\bibitem{raghavan2007near}
U.~N. Raghavan, R.~Albert, and S.~Kumara, ``Near linear time algorithm to
  detect community structures in large-scale networks,'' \emph{Physical review
  E}, vol.~76, no.~3, p. 036106, 2007.

\bibitem{shiokawa2013fast}
H.~Shiokawa, Y.~Fujiwara, and M.~Onizuka, ``Fast algorithm for modularity-based
  graph clustering,'' in \emph{Twenty-Seventh AAAI Conference on Artificial
  Intelligence}, 2013.

\bibitem{powerlyra}
R.~Chen, J.~Shi, Y.~Chen, and H.~Chen, ``Powerlyra: Differentiated graph
  computation and partitioning on skewed graphs,'' in \emph{Proceedings of the
  Tenth European Conference on Computer Systems}, ser. EuroSys '15.\hskip 1em
  plus 0.5em minus 0.4em\relax New York, NY, USA: ACM, 2015, pp. 1:1--1:15.

\bibitem{199295}
X.~Zhu, W.~Chen, W.~Zheng, and X.~Ma, ``Gemini: A computation-centric
  distributed graph processing system,'' in \emph{12th {USENIX} Symposium on
  Operating Systems Design and Implementation ({OSDI} 16)}.\hskip 1em plus
  0.5em minus 0.4em\relax Savannah, GA: {USENIX} Association, Nov. 2016, pp.
  301--316.

\bibitem{graphchi}
A.~Kyrola, G.~Blelloch, and C.~Guestrin, ``Graphchi: Large-scale graph
  computation on just a {PC},'' in \emph{Presented as part of the 10th {USENIX}
  Symposium on Operating Systems Design and Implementation ({OSDI} 12)}.\hskip
  1em plus 0.5em minus 0.4em\relax Hollywood, CA: {USENIX}, 2012, pp. 31--46.

\bibitem{gridgraph}
X.~Zhu, W.~Han, and W.~Chen, ``Gridgraph: Large-scale graph processing on a
  single machine using 2-level hierarchical partitioning,'' in \emph{2015
  {USENIX} Annual Technical Conference ({USENIX} {ATC} 15)}.\hskip 1em plus
  0.5em minus 0.4em\relax Santa Clara, CA: {USENIX} Association, Jul. 2015, pp.
  375--386.

\bibitem{Maass:2017:MPT:3064176.3064191}
S.~Maass, C.~Min, S.~Kashyap, W.~Kang, M.~Kumar, and T.~Kim, ``Mosaic:
  Processing a trillion-edge graph on a single machine,'' in \emph{Proceedings
  of the Twelfth European Conference on Computer Systems}, ser. EuroSys
  '17.\hskip 1em plus 0.5em minus 0.4em\relax New York, NY, USA: ACM, 2017, pp.
  527--543.

\bibitem{Alistarh:2015:SMH:2969442.2969452}
D.~Alistarh, J.~Iglesias, and M.~Vojnovic, ``Streaming min-max hypergraph
  partitioning,'' in \emph{Proceedings of the 28th International Conference on
  Neural Information Processing Systems - Volume 2}, ser. NIPS'15.\hskip 1em
  plus 0.5em minus 0.4em\relax Cambridge, MA, USA: MIT Press, 2015, pp.
  1900--1908.

\bibitem{1639359}
K.~D. Devine, E.~G. Boman, R.~T. Heaphy, R.~H. Bisseling, and U.~V. Catalyurek,
  ``Parallel hypergraph partitioning for scientific computing,'' in
  \emph{Proceedings 20th IEEE International Parallel Distributed Processing
  Symposium}, April 2006, pp. 10 pp.--.

\bibitem{TRIFUNOVIC2008563}
A.~Trifunović and W.~J. Knottenbelt, ``Parallel multilevel algorithms for
  hypergraph partitioning,'' \emph{Journal of Parallel and Distributed
  Computing}, vol.~68, no.~5, pp. 563 -- 581, 2008.

\bibitem{Kabiljo:2017:SHP:3137628.3137650}
I.~Kabiljo, B.~Karrer, M.~Pundir, S.~Pupyrev, and A.~Shalita, ``Social hash
  partitioner: A scalable distributed hypergraph partitioner,'' \emph{Proc.
  VLDB Endow.}, vol.~10, no.~11, pp. 1418--1429, Aug. 2017.

\bibitem{8621968}
C.~Mayer, R.~Mayer, S.~Bhowmik, L.~Epple, and K.~Rothermel, ``Hype: Massive
  hypergraph partitioning with neighborhood expansion,'' in \emph{2018 IEEE
  International Conference on Big Data (Big Data)}, Dec 2018, pp. 458--467.

\end{thebibliography}

\end{document}